# Cooperative Games with Overlapping Coalitions

**Georgios Chalkiadakis**                                    GC2@ECS.SOTON.AC.UK
*School of Electronics and Computer Science,*
*University of Southampton, SO17 1BJ, UK*

**Edith Elkind**                                             EELKIND@NTU.EDU.SG
*School of Physical and Mathematical Sciences,*
*Nanyang Technological University, 637371, Singapore*

**Evangelos Markakis**                                       MARKAKIS@GMAIL.COM
*Department of Infomatics,*
*Athens University of Economics and Business, GR10434, Greece*

**Maria Polukarov**                                          MP3@ECS.SOTON.AC.UK
**Nicholas R. Jennings**                                     NRJ@ECS.SOTON.AC.UK
*School of Electronics and Computer Science,*
*University of Southampton, SO17 1BJ, UK*

## Abstract

In the usual models of cooperative game theory, the outcome of a coalition formation process is either the grand coalition or a coalition structure that consists of disjoint coalitions. However, in many domains where coalitions are associated with tasks, an agent may be involved in executing more than one task, and thus may distribute his resources among several coalitions. To tackle such scenarios, we introduce a model for *cooperative games with overlapping coalitions*—or *overlapping coalition formation (OCF) games*. We then explore the issue of stability in this setting. In particular, we introduce a notion of the core, which generalizes the corresponding notion in the traditional (non-overlapping) scenario. Then, under some quite general conditions, we characterize the elements of the core, and show that any element of the core maximizes the social welfare. We also introduce a concept of balancedness for overlapping coalitional games, and use it to characterize coalition structures that can be extended to elements of the core. Finally, we generalize the notion of convexity to our setting, and show that under some natural assumptions convex games have a non-empty core. Moreover, we introduce two alternative notions of stability in OCF that allow a wider range of deviations, and explore the relationships among the corresponding definitions of the core, as well as the classic (non-overlapping) core and the Aubin core. We illustrate the general properties of the three cores, and also study them from a computational perspective, thus obtaining additional insights into their fundamental structure.

## 1. Introduction

*Coalition formation*, widely studied in game theory and economics (Myerson, 1991), has attracted much attention in AI as means of forming teams of autonomous selfish agents that need to cooperate to perform certain tasks (Sandholm & Lesser, 1997; Shehory & Kraus, 1998; Sandholm, Larson, Andersson, Shehory, & Tohme, 1999; Manisterski, Sarne, & Kraus, 2008; Rahwan, Ramchurn, Jennings, & Giovannucci, 2009). Traditionally, in the game theory literature it is assumed that the outcome of the coalition formation process is either the *grand coalition* (i.e., the set of all agents), or a *coalition structure* that consists of disjoint coalitions (i.e., *a partition* of the set of agents). While natural for some settings, in many scenarios of interest this assumption is not applicable.

Specifically, it is often natural to associate coalitions with tasks to be performed by the agents. In such situations, some agents may be involved in several tasks, and therefore may need to distribute their resources among the coalitions in which they participate. Indeed, such "overlaps" may be necessary to obtain a good outcome, and are natural in a plethora of interesting applications. As a simple e-commerce example, consider online trading agents representing individuals or virtual enterprises, and facing the challenge of allocating their owners' capital to a variety of projects (i.e., coalitions) simultaneously. There are many other examples of settings in which an agent (be it a software entity or a human) splits his resources (such as processing power, time or money) among several tasks. These tasks, in turn, may require the participation of more than one agent: a computation may run on several servers, a software project usually involves more than one engineer, and a start-up may rely on several investors. Thus, each task corresponds to a coalition of agents, but agents' contributions to those coalitions may be fractional, and, moreover, agents can participate in several tasks at once, resulting in *coalition structures with overlapping coalitions*. The formation of overlapping coalitions is particularly prevalent in systems demanding multiagent or multirobot coordination, computational grid networks, and sensor networks—see, e.g., the work of Patel et al. (2005), and Dang, Dash, Rogers, & Jennings (2006). To date, however, there has been essentially no theoretical treatment of the topic, with just a few exceptions (which we discuss in Section 3).

Against this background, the goal of this paper is to introduce and study a model that explicitly takes *overlapping coalition formation (OCF)* into account. Our model is applicable in situations where agents need to allocate different parts of their resources to simultaneously serve different tasks as members of different coalitions. Besides allowing for overlapping coalitions, it departs from the conventional coalition formation framework in two important aspects. First, there is no inherent superadditivity assumption in our work, and hence the grand coalition does not always emerge. Thus, our subsequent definition of the core incorporates coalition structures. Second, exactly because we are interested in outcomes other than the grand coalition formation, we do not use the standard *transferable utility (TU)* framework, where agents can make arbitrary payments to each other. Instead, following the seminal paper by Aumann and Dreze (1974), we allow arbitrary monetary transfers *within* coalitions, but not cross-coalitional transfers. That is, an agent not contributing to a coalition should not expect to receive payoff from it. Indeed, as argued by Aumann and Dreze, the inability of some of the agents to work together and share payoffs may be one of the primary reasons why the grand coalition does not form, and a particular coalition structure arises. Finally, our model can take task (coalitional action) execution explicitly into account; this facilitates possible extensions to tackle coalition formation under uncertainty.[1]

Apart from defining a model for overlapping coalition formation, the main contribution of this work is exploring the stability concept of the *core* in the OCF setting. We suggest three different notions of the core, depending on the nature of deviations allowed, since, as we shall see, the range of permissible deviations in an overlapping setting can be much richer than in the traditional non-overlapping one. More specifically, the definition of stability depends on whether a deviator who reduced his contribution to some—but not all—coalitions, expects to get any payoff from the coalitions that he did not abandon completely.

To provide more intuition, consider the example of two construction companies, 1 and 2, who are currently partners (not necessarily the only partners) working on construction projects A ("building a university campus") and B ("building a hospital"). Assume that partner 1 has more stakes in project B, expecting to extract from it a great value, and has contributed to it 75% of its available resources, contributing the remaining 25% to A; while partner 2 contributes most of its resources (say 67%) to project A and the remaining fraction (say 33%) to B. Thus, they currently participate in two overlapping coalitions, each one performing a different task. Now, if partner 2 feels unhappy about the current payoff division arrangement, it might consider abandoning project A (by cancelling the project if it is the project leader, or by taking advantage of some contractual exit clause) in order to commit its resources to a more profitable to 2 project (say C). However, by doing so, it might hurt project A's chances of completion. Does this mean that 2's actions will trigger the spite of company 1, which might use available means to kick 2 out of project B? And what if company 2 lowered its degree of participation in A instead of withdrawing completely? How much of the profits from completing A would 2 then be entitled to? The different answers one can provide to these questions correspond to different notions of profitable deviations, and, therefore, to different notions of core-stability. In particular, we demonstrate that the core notions we put forward in this paper are substantially different from each other with respect to the sets of outcomes they characterize.

---

1. To simplify notation, we only show how to incorporate coalitional actions in the model in Section 10.

Our main technical results involve the *c-core*, the first core concept that we suggest. Among the three concepts of the core introduced in this paper, the c-core is the closest to the standard definition of the core in general non-transferable utility (NTU) games. In particular, we provide conditions for the existence of the c-core as follows. Under quite general assumptions, we first provide a characterization for outcomes, i.e., pairs of the form *(overlapping coalition structure, imputation)*, to be in the c-core. Our proof is based on a graph-theoretic argument, which may be of independent interest. As a corollary of this result, we show that any outcome in the c-core maximizes the social welfare. Second, we characterize coalition structures that admit payoff allocations such that the resulting outcome is in the c-core. This is done by generalizing the Bondareva-Shapley theorem to our setting (note that this theorem does not hold for arbitrary non-transferable utility games). Furthermore, we extend the notion of convexity in coalitional games to overlapping coalitions, and show that under mild assumptions any convex OCF game has a non-empty c-core.

We then discuss the properties of all three versions of the OCF-core we suggest, and relate them to each other and to the classic core. We also demonstrate how our model and core concepts differ from fuzzy coalitional games (Aubin, 1981); though relevant to that model, our work is fundamentally different. In addition, we initiate the study of computational aspects of stability in the overlapping setting. Note that the computational analysis of coalitional games, even in non-overlapping scenarios, is hindered by the fact that, in general, coalitional games do not possess a compact representation, as one may have to list the value of every possible coalition. Thus, the existing work on algorithmic aspects of coalitional games focused on game representations that are either incomplete—such as, e.g., weighted voting games (Elkind, Goldberg, Goldberg, & Wooldridge, 2009), induced subgraph games (Deng & Papadimitriou, 1994), or network flow games (Bachrach & Rosenschein, 2007)—or are only guaranteed to be succinct for specific subclasses of games, such as MC-nets (Ieong & Shoham, 2005) or coalitional skill games (Bachrach & Rosenschein, 2008); another approach is to show complexity bounds for all games representable by polynomial-sized circuits (Greco, Malizia, Palopoli, & Scarcello, 2009). This issue is even more severe in the OCF setting, as now we have to specify the value of every *partial* coalition. Therefore, in this paper, we follow the first of these approaches, and introduce a formalism of *threshold task games* that is capable of describing a large family of overlapping coalition formation settings in a succinct manner. Within this formalism, we obtain both negative and positive results regarding the complexity of deciding the questions of membership and non-emptiness for our OCF-core concepts. We conclude by describing some natural extensions of our model and suggesting directions for future work.[2]

## 2. Preliminaries

In this section, we provide a brief overview of the basic concepts in cooperative game theory regarding non-overlapping coalition structures. To begin, let $N = \{1, \ldots, n\}$ be a set of players (or "agents"). A subset $S \subseteq N$ is called a *coalition*. A *coalition structure* $(CS)$ in non-overlapping environments is a partition of the set of agents.

Under the assumption of *transferable utility*, coalition formation can be abstracted into a fairly simple model. This assumption postulates the existence of a (divisible) commodity (e.g., "money") that can be freely transferred among players. The role of the *characteristic function* of a *coalitional game with transferable utility (TU-game)* is to specify a single number denoting the worth of a coalition. Formally, a characteristic function $v : 2^N \mapsto \mathbb{R}$ defines the *value* $v(S)$ of each coalition $S$ (von Neumann & Morgenstern, 1944). A transferable utility game $G$ is completely specified by the set of players $N$ and the characteristic function $v$; we can therefore write $G = (N, v)$.

While the characteristic function describes the payoffs available to coalitions, it does not prescribe a way of distributing these payoffs. This is captured by the notion of an *imputation*, defined as follows. We say that an *allocation* is a vector of payoffs $\boldsymbol{x} = (x_1, \ldots, x_n)$ assigning some payoff to each $j \in N$. An allocation $\boldsymbol{x}$ is *efficient* with respect to a coalition structure $CS$ if $\sum_{j \in S} x_j = v(S)$ for all $S \in CS$; and it is called an *imputation* if it is efficient and satisfies *individual rationality*, i.e., $x_j \geq v(\{j\})$ for $j = 1, \ldots, n$. The set of all imputations of $CS$ is denoted by $I(CS)$.

Now, when rational agents seek to maximize their individual payoffs, the *stability* of the underlying coalition structure becomes critical, as agents might be tempted to abandon agreements in pursuit of further gains for themselves. A structure is stable only if the outcomes attained by the coalitions and the payoff combinations agreed to by the agents satisfy both individual and group rationality. Given this requirement, research in coalition formation has developed several notions of stability, among the strongest and the most well-studied ones being the *core* (Gillies, 1953). Taking coalition structures into account, the core of a TU game is a set of outcomes $(CS, \boldsymbol{x})$, $\boldsymbol{x} \in I(CS)$, such that no subgroup of agents is motivated to depart from their coalitions in $CS$.

**Definition 1.** *Let $CS$ be a coalition structure, and let $\boldsymbol{x} \in \mathbb{R}^n$ be an allocation of payoffs to the agents. The* core *of a TU game $(N, v)$ is the set of all pairs $(CS, \boldsymbol{x})$ such that $\boldsymbol{x} \in I(CS)$ and for any $S \subseteq N$ it holds that $\sum_{j \in S} x_j \geq v(S)$.*

Hence, no coalition would ever "block" the proposal for a core allocation. It is well-known that the core is a strong notion, and there exist many games where it is empty (Myerson, 1991).

The core definition above is essentially the definition provided by Sandholm and Lesser (1997) (and is also very similar to the one given by Dieckmann & Schwalbe, 1998). If we assume super-additivity of the characteristic function (i.e., $v(U \cup T) \geq v(U) + v(T)$ for any disjoint coalitions $U$ and $T$) then in the definition above we may only consider outcomes where $CS$ is simply the grand

---

coalition and $\sum_{j \in N} x_j = v(N)$. The core definition then becomes the traditional definition that has been used in the vast majority of the economics literature (Osborne & Rubinstein, 1994).

The environments of interest in our work however are mainly non-superadditive and we will not make any such assumption on the characteristic function. Indeed, there is a plethora of realistic application scenarios where the emergence of the grand coalition is either not guaranteed, might be perceivably harmful, or is plainly impossible (Sandholm & Lesser, 1997; Sandholm et al., 1999). In addition to such motivations, Aumann and Dreze (1974) also provide a thorough and insightful discussion on why coalition structures arise: they put forward a series of arguments on how this might happen, and explain that coalition structures may emerge naturally even in superadditive environments for a variety of reasons. Briefly, their arguments describe how a subset of agents might find it more worthwhile to bargain within the framework of a specific structure, than within the framework of the grand coalition; or how the emergence of a coalition structure may reflect considerations that are by necessity excluded from the formal description of the game because they are impossible to measure or communicate. Exogenous arguments for the emergence of coalition structures naturally include the impossibility of communication among all negotiators, or the *by law* prohibition of the grand coalition (Aumann & Dreze, 1974).

## 3. Related Work

The work that is most relevant to ours is the research on *fuzzy coalitional games*, introduced by Aubin (1981). Branzei, Dimitrov, & Tijs (2005) also provide a detailed exposition of such games. A player in a fuzzy game can participate in a coalition at various *levels*, and the value of a coalition $S$ depends on the participation levels of the agents in $S$. Given this model, Aubin then defines the core for fuzzy games (also referred to as the *Aubin core*). Though our model also allows for partial participation in a coalition, there are several crucial differences between fuzzy games and OCF games, and the corresponding notions of stability. We postpone listing these until after presenting our model and results, but will do so in Section 8.2. For now, let us just point out that, in distinction to our work, the formation of coalition structures (overlapping or not) is not addressed in the fuzzy games literature.

Apart from fuzzy games, very little work exists on overlapping coalition formation settings. Here we discuss some notable exceptions, as well as some related work on the core in the context of non-overlapping coalition structures.

To begin, Shehory and Kraus (1996) present a setting for overlapping coalition formation. In their model, the agents have goals and capabilities—i.e., abilities to execute certain actions. To serve their goals, the agents have to participate in coalitions, to each of which they contribute some of their capabilities, which can thus be thought of as resources. The authors then propose heuristic algorithms that lead to the creation of overlapping coalition structures. However, the authors stop short of addressing the question of the stability of overlapping coalitions. Dang et al. (2006) also examine heuristic algorithms for overlapping coalition formation to be used in surveillance multi-sensor networks. However, their work does not deal with payoff allocation issues, and does not view the overlapping coalition formation problem from a game-theoretic perspective.

Conconi and Perroni (2001) present a model of international multidimensional policy coordination in a *non-cooperative* setting: agreement structures between countries can be overlapping, namely a country may participate in multiple agreements, by contributing any number of proposed "elementary strategies" (which can be regarded as being chosen from *discrete* sets of resources).

They then introduce an equilibrium concept to describe stability in this setting. However, in contrast to our work, the setting in the work of Conconi and Perroni is non-cooperative, and does not apply to agents with continuous resources.

More recently, Albizuri, Aurrecoechea, & Zarzuelo (2006) presented an extension of Owen's value (1977)—which, in turn, can be thought of as a generalization of the Shapley value (1953)—to an overlapping coalition formation setting. Specifically, they present an axiomatic characterization of their *configuration value*. However, in the work of Albizuri et al. there exists no notion of resources that an agent needs to distribute across coalitions.

With regard to non-overlapping coalition structures as presented in Section 2, Sandholm and Lesser (1997) examine the problem of allocating *computational resources* to coalitions. They do not restrict themselves to superadditive settings, but discuss the stability of coalition structures instead. In particular, they introduce a notion of bounded rational core that explicitly takes into account coalition structures. Apt & Radzik (2006) and Apt & Witzel (2009) also do not restrain themselves to coalition formation problems where the outcome is the grand coalition only. Instead, they introduce various stability notions for abstract games whose outcomes can be coalition structures, and discuss simple transformations (e.g., split and merge rules) by which stable partitions of the set of players may emerge. However, none of these papers considers any extensions to overlapping coalitions.

## 4. Our Model

In this section we extend the traditional model of Section 2 to cooperative games with overlapping coalitions. In most scenarios of interest, even if overlapping coalitions are allowed, an agent would not be able to participate in all possible coalitions due to lack of time, cash flow, or energy. To model this, we assume that each agent possesses a certain amount of resources which he can distribute among the coalitions he joins. Without loss of generality, we can make a normalization and assume that each agent has one unit of resource: an agent's contribution to a coalition is thus given by the fraction of his resources that he allocates to it. We can also think of this as the agent's "participation level", or the fraction of time he devotes to a coalition. Of course, an agent may own several types of resources (e.g., time *and* money), and his contribution to a coalition would then be described by a vector rather than a scalar. Our model, and all of our results, extend to this more general setting in a straightforward manner. Nevertheless, for conciseness, we restrict our presentation to the single-resource setting.

As discussed above, in the non-overlapping model a coalition is a subset of agents, and a game is defined by its characteristic function $v : 2^N \mapsto \mathbb{R}$, representing the maximum total payoff that a coalition can get. In our setting, a *partial coalition* is given by a vector $\boldsymbol{r} = (r_1, \ldots, r_n)$, where $r_j$ is the fraction of agent $j$'s resources contributed to this coalition ($r_j = 0$ means that $j$ is not a member of the coalition). The *support* of a partial coalition $\boldsymbol{r}$ is denoted by $\text{supp}(\boldsymbol{r})$ and is defined as $\text{supp}(\boldsymbol{r}) = \{j \in N \mid r_j \neq 0\}$. We can now define the *cooperative games with overlapping coalitions*, or *overlapping coalition formation games* (OCF-games for short), which we will be considering in the rest of this work.

**Definition 2.** *An OCF-game $G$ with player set $N = \{1, \ldots, n\}$ is given by a function $v : [0, 1]^n \to \mathbb{R}$, where $v(0^n) = 0$.*

Function $v$ maps each partial coalition $\boldsymbol{r}$ to the corresponding payoff. We denote this game by $G = (N, v)$, or, if $N$ is clear from the context, simply by $v$. Clearly, a "classic" coalition $S \subseteq N$ can now be represented as the vector $e^S$, where $(e^S)_j = 1$ for $j \in S$ and 0 otherwise. In the economics literature, these are sometimes called *crisp* coalitions, whereas coalitions of the form $(r_1, \ldots, r_n)$ with at least one $r_j$ in $(0, 1)$ are referred to as *fuzzy* coalitions (Branzei et al., 2005). We will avoid the latter term in our work so as not to cause confusion with fuzzy games, and refer instead to coalitions of this kind as *partial* coalitions, or simply coalitions.

In most scenarios of interest, $v$ is *monotone*, i.e., satisfies $v(\boldsymbol{r}) \geq v(\boldsymbol{r'})$ for any $\boldsymbol{r}, \boldsymbol{r'}$ such that $r_j \geq r'_j$ for all $j = 1, \ldots, n$. Note that if $v$ is monotone, we have $v(\boldsymbol{r}) \geq 0$ for any $\boldsymbol{r} \in [0, 1]^n$, since we set $v(0, \ldots, 0) = 0$. In our discussion of stability of overlapping coalitions, we will assume that $v$ is monotone.

We now need to specify the possible outcomes of an OCF-game. In the non-overlapping setting, an outcome is a pair $(CS, \boldsymbol{x})$, where $CS$ is a partition on $N$ and $\boldsymbol{x}$ is an imputation for $CS$. To extend this definition to our scenario, we start by introducing the notion of a coalition structure with overlapping coalitions. While we will be mostly interested in coalition structures over $N$, the definition below is given for coalition structures over an arbitrary subset $T \subseteq N$, as this will be useful for defining the maximum profit a subset of agents can achieve (see the definition of the function $v^*$ below).

**Definition 3.** *For a set of agents $T \subseteq N$, a coalition structure on $T$ is a finite list of vectors (partial coalitions) $CS_T = (\boldsymbol{r}^1, \ldots, \boldsymbol{r}^k)$ that satisfies (i) $\boldsymbol{r}^i \in [0, 1]^n$; (ii) $\operatorname{supp}(\boldsymbol{r}^i) \subseteq T$ for all $i = 1, \ldots, k$; and (iii) $\sum_{i=1}^k r_j^i \leq 1$ for all $j \in T$. We will refer to $k$ as the* size *of the coalition structure $CS_T$ and write $|CS_T| = k$. Also, $\mathcal{CS}_T$ denotes the set of all coalition structures on $T$.*

In the definition above, each $\boldsymbol{r}^i = (r_1^i, r_2^i, \ldots, r_n^i)$ corresponds to some partial coalition ($r_j^i$ being the fraction of the resources that agent $j$ contributes to $\boldsymbol{r}^i$). The constraints state that every agent from $T$ distributes at most one unit of his resources among the various coalitions he participates in (those may include the singleton coalition). This allows coalitions to be overlapping. Note that the coalition structure is a list rather than a set, i.e., it can contain two or more identical partial coalitions. Observe also that an agent is not required to allocate all of his resources, i.e., it can be the case that $\sum_{i=1}^k r_j^i < 1$. However, under monotonicity, we can assume that for each agent $j$ we have $\sum_{i=1}^k r_j^i = 1$ (i.e., a coalition structure is a fractional partition of the agents).

We would like to remark that one could conceive of other models that also allow agents to form overlapping coalitions. As an example, instead of requiring agents to distribute at most one unit of resources among partial coalitions, we could have constraints on the number of (crisp) coalitions an agent could take part in. While we believe that our model is flexible enough to represent a wide range of realisitc scenarios, and we focus on it throughout our work, in Section 10, we discuss several extensions of our model.

The introduction of overlapping coalition structures imposes some new technical challenges. For instance, while in the non-overlapping setting the number of different coalition structures is finite, in our setting there can be infinitely many different partial coalitions, and hence infinitely many coalition structures. This implies that it is impossible to find the social welfare-maximizing coalition structure by enumerating all candidate solutions—in fact, the maximum may not even be attained. In contrast, in a non-OCF setting this approach is possible—though, in general, infeasible.

We now extend the definition of $v$ to coalition structures by setting $v(CS) = \sum_{\boldsymbol{r} \in CS} v(\boldsymbol{r})$. Furthermore, for any $S \subseteq N$ we define $v^*(S) = \sup_{CS \in \mathcal{CS}_S} v(CS)$. Intuitively, $v^*(S)$ is the least upper bound on the value that the members of $S$ can achieve by forming a coalition structure; for the interested reader, we note that it corresponds to the characteristic function of the game's *superadditive cover* (Aumann & Dreze, 1974). Clearly, $v^*(S)$ may exceed the value of coalition $S$ itself, i.e., $v(e^S)$, since it may be profitable for the players in $S$ to form several overlapping coalitions over $S$. We say that $v$ is *bounded* if $v^*(N) < \infty$; for most games of interest, $v$ is likely to be bounded.

As in our setting the agents will not necessarily form the grand coalition, we will be interested in reasoning about coalition structures from $\mathcal{CS}_N$. The coalition structure will impose restrictions on admissible ways of distributing the gains; a payoff vector corresponds to an imputation if and only if it is obtained by distributing the value of each coalition:

**Definition 4.** *Given a coalition structure $CS \in \mathcal{CS}_N$, $|CS| = k$, an* imputation *for $CS$ is a $k$-tuple* $\boldsymbol{x} = (\boldsymbol{x}^1, \ldots, \boldsymbol{x}^k)$*, where $\boldsymbol{x}^i \in \mathbb{R}^n$ for $i = 1, \ldots, k$, such that*

- *(Payoff Distribution) for every partial coalition $\boldsymbol{r}^i \in CS$ we have $\sum_{j=1}^n x_j^i = v(\boldsymbol{r}^i)$ and $r_j^i = 0$ implies $x_j^i = 0$;*

- *(Individual Rationality) the total payoff of agent $j$ is at least as large as what he can achieve on his own: $\sum_{i=1}^k x_j^i \geq v^*(\{j\})$.*

The set of all imputations for $CS$ is denoted by $I(CS)$. Notice that in Definition 4, the profit from a task assigned to a partial coalition is only distributed among agents involved in executing it. Thus, no transfers of that payoff are allowed to outsiders. Note also that the individual rationality constraint is defined in terms of $v^*$ rather than $v$, as even for a single agent it may be profitable to split into several partial coalitions (e.g., if there are many tasks, each of which only requires a small fraction of his resources).

Now, the set of outcomes that is of interest to us is the set of *feasible agreements*:

**Definition 5.** *A feasible agreement (or an* outcome*) for a set of agents $J \subseteq N$ is a tuple $(CS, \boldsymbol{x})$ where $CS \in \mathcal{CS}_J$, $|CS| = k$ for some $k \in \mathbb{N}$, and $\boldsymbol{x} = (\boldsymbol{x}^1, \ldots, \boldsymbol{x}^k) \in I(CS)$. We denote the set of all feasible agreements for $J$ by $\mathcal{F}(J)$.*

The payoff $p_j$ of an agent $j$ under a feasible agreement $(CS, \boldsymbol{x})$ is $p_j(CS, \boldsymbol{x}) = \sum_{i=1}^k x_j^i$. We write $\boldsymbol{p}(CS, \boldsymbol{x})$ to denote the vector $(p_1(CS, \boldsymbol{x}), \ldots, p_n(CS, \boldsymbol{x}))$. Finally, note that it is straightforward to extend the definitions above to games on subsets of the agents. In particular, we require that an imputation $\boldsymbol{x} \in I(CS_J)$ satisfies $x_j^i = 0$ for $j \notin J$.

Given this model, we are now ready to define the concept of the core for cooperative games with overlapping coalitions.

## 5. The Core with Overlapping Coalitions

In this section, we investigate several approaches to defining stability in OCF-games. Specifically, here we propose and analyze three alternative definitions of the core.

Before presenting the core definitions, we define a new class of games, which we will be using as our running example, namely the class of *threshold task games* (TTGs). TTGs form a simple, but expressive class of coalitional games, and can be used to model collaboration in multi-agent systems. In TTGs agents pool resources in order to accomplish tasks, so the idea of agents contributing resources to more than one task and thus participating in several coalitions simultaneously is extremely natural in this context. Thus, and due to their simplicity, TTGs provide a convenient vehicle for the study of core-stability in the overlapping setting, and we will be using them for this purpose throughout the rest of the paper (though our work is not limited to this class of games).

### 5.1 Threshold Task Games

Threshold task games are defined as follows.

**Definition 6.** *A threshold task game* $G = (N; \boldsymbol{w}; \boldsymbol{t})$ *is given by:*

- *a set of agents* $N = \{1, \ldots, n\}$;

- *a vector* $\boldsymbol{w} = (w_1, \ldots, w_n) \in \mathbb{R}^+$ *of the agents' weights*;

- *a list* $\boldsymbol{t} = (t^1, \ldots, t^m)$ *of task types, where each task type* $t^j$ *is described by a threshold* $T^j \geq 0$ *and a utility* $u^j \geq 0$; *we write* $t^j = (T^j, u^j)$.

Intuitively, such games describe scenarios where agents can split into teams to work on tasks. There is one type of resource (e.g., time or money) that is needed for all tasks, and each agent has a certain amount of this resource which corresponds to his weight $w_i$ (we chose the term "weight" to avoid confusion with the use of the term "resource" in the context of OCF-games). There are $m$ types of tasks, each of which is described by a resource requirement $T^j$ and a utility $u^j$. If the team of agents that works on $t^j$ has total weight at least $T^j$, this means that it has sufficient resources to complete the task, so it obtains the full value of this task $u^j$. Otherwise, its payoff from this task is 0. We assume that there are infinitely many tasks of each type, so that if one team of agents chooses to work on $t^j$, this does not prevent another team from choosing $t^j$ as well. In what follows, we assume that the list $\boldsymbol{t}$ is *monotone*, i.e., it satisfies $T^1 < \ldots < T^m$ and $u^1 < \ldots < u^m$. Indeed, if there are two task types $t^i, t^j$ such that $T^i \leq T^j$, but $u^i \geq u^j$, we can safely assume that no team of agents will choose to work on $t^j$, and hence $t^j$ can be deleted from $\boldsymbol{t}$. Hence, our monotonicity assumption can be made without loss of generality.

The description above suggests that we can interpret a TTG $G = (N, \boldsymbol{w}, \boldsymbol{t})$ as a (non-overlapping) coalitional game $\hat{G} = (N, \hat{v})$, where for $S \subseteq N$ we set

$$\hat{v}(S) = \max\{0, \max\{u^j \mid w(S) \geq T^j\}\}$$

(note that we use the standard convention $\max \emptyset = -\infty$). Such games provide a direct generalization of weighted voting games (WVGs) with coalition structures introduced by Elkind, Chalkiadakis, & Jennings (2008). Indeed, WVGs with coalition structures can be seen as TTGs in which there is only one task type $t = t^1$ with utility 1.

At the same time, one can also interpret TTGs as games with overlapping coalitions by allowing each agent to spread his weight across several tasks. The corresponding OCF-game $\check{G} = (N, \check{v})$ is given by

$$\check{v}(r_1, \ldots, r_n) = \max\{0, \max\{u^j \mid \sum_{i=1}^{n} r_i w_i \geq T^j\}\}.$$

That is, a partial coalition can successfully complete a task of type $t^j$ and earn its value $u^j$ if the total weight contributed by all agents to this partial coalition is at least $T^j$.

**Example 1.** *Consider a TTG* $G = (N; \boldsymbol{w}; \boldsymbol{t})$, *where* $N = \{1, 2, 3\}$, $\boldsymbol{w} = (2, 2, 2)$ *and* $\boldsymbol{t} = t^1 = (3, 1)$. *For the corresponding non-overlapping game* $\hat{G}$ *we have* $\hat{v}(\{1\}) = 0$, $\hat{v}(\{1, 2\}) = \hat{v}(\{1, 2, 3\}) = 1$. *Note that when overlapping coalitions are not allowed, the maximum social welfare achievable by any coalition structure over* $N$ *is 1, as agents cannot split into two disjoint groups each of which having weight at least 3.*

*In contrast, for the corresponding OCF-game $\check{G} = (N, \check{v})$ we have $\check{v}(1, 0, 0) = 0$, $\check{v}(1, 1, 0) = \check{v}(1, 1, 1) = 1$, and, moreover, $\check{v}(1, .5, 0) = 1$ and $\check{v}(0, .5, 1) = 1$. Hence the maximum social welfare is 2 in the overlapping setting since the second agent can split his weight between two coalitions so that each of them has enough resources to complete the task.*

From Example 1, it should be clear that for any TTG $G$, the maximum social welfare achievable in its overlapping version $\hat{G}$ is at least as large as the maximum social welfare in its non-overlapping version $\hat{G}$—i.e., allowing agents to split their weights between the tasks can only increase efficiency. Moreover, this increase can be arbitrarily large even for a single agent. Indeed, consider one agent of weight $w$ and one task type $t$ with $T = 1$, $u = 1$. If overlapping coalitions are not allowed, the agent's total utility is 1, while in the overlapping scenario he can obtain $w$. For the interested reader, Appendix A discusses algorithmic aspects of social welfare maximization in TTGs, both in the overlapping and in the non-overlapping scenario.

## 5.2 Three Definitions of the Core

As explained in Section 2 above, core-stability implies that no group of agents should be able to profitably deviate from a configuration in the core. Hence, any definition of the core has to depend on the notion of permissible deviations used. Now, in the non-overlapping setting a deviator abandons the coalition he originally participated in, and joins a new coalition. Thus, there is no reason why he should obtain any payoff from the coalition that he left. In the overlapping setting, the situation is less clear-cut. Indeed, when deviating, an agent may abandon some coalitions completely, withdraw some—but not all—of his contribution to other coalitions, and keep his contribution to the remaining coalitions unchanged. The question then is whether this agent should expect to obtain any payoff from the partial coalitions with non-deviators that he is still contributing to.

Our first notion of the core assumes that the answer to this question is "no". Thus, once an agent is identified as a deviator—i.e., he alters his contribution to any given coalition—he no longer expects to benefit from his cooperation with non-deviators. By monotonicity, this means that the deviators have nothing to gain from contributing resources to coalitions with non-deviators. Therefore, under the first definition of the core which we present here, we assume that the deviators only form coalitions among themselves, or, in other words, each deviation can be seen as an overlapping coalition structure over the set of deviators. We remark that this definition can be seen as the most straightforward generalization of the standard notion of the core: indeed, just as in the standard setting, each deviator completely withdraws from coalitions with non-deviators, and only benefits from coalitions with other deviators. We formalize this approach as follows.

**Definition 7.** *Given an OCF-game $G = (N, v)$ and a set of agents $J \subseteq N$, let $(CS, \boldsymbol{x})$ and $(CS', \boldsymbol{y})$ be two outcomes of $G$ such that for any partial coalition $\boldsymbol{s}^\ell \in CS'$ either $\mathrm{supp}(\boldsymbol{s}^\ell) \subseteq J$ or $\mathrm{supp}(\boldsymbol{s}^\ell) \subseteq N \setminus J$. Then we say that $(CS', \boldsymbol{y})$ is a* profitable deviation *of $J$ from $(CS, \boldsymbol{x})$ if for all $j \in J$ we have $p_j(CS', \boldsymbol{y}) > p_j(CS, \boldsymbol{x})$. We say that an outcome $(CS, \boldsymbol{x})$ is in the* core *of $G$ if no subset of agents $J$ has a profitable deviation from it. That is, for any set of agents $J \subseteq N$, any coalition structure $CS_J$ on $J$, and any imputation $\boldsymbol{y} \in I(CS_J)$, we have $p_j(CS_J, \boldsymbol{y}) \leq p_j(CS, \boldsymbol{x})$ for some agent $j \in J$.*

In this definition, the deviation $CS'$ is restricted to be a coalition structure in which there are no partial coalitions involving both the deviators and the non-deviators—i.e., each partial coalition contains either deviators only ($\mathrm{supp}(\boldsymbol{s}^\ell) \subseteq J$) or non-deviators only ($\mathrm{supp}(\boldsymbol{s}^\ell) \subseteq N \setminus J$). Thus, any payoff that the players in $J$ can receive under $CS'$ would have to come from partial coalitions over $J$ only.

**Example 2.** *Consider the OCF-game $\check{G}$ that corresponds to a threshold task game $G = (N; \boldsymbol{w}; \boldsymbol{t})$, where $N = \{1, 2\}$, $\boldsymbol{w} = (4, 6)$, and $\boldsymbol{t} = (t^1, t^2)$ with $t^1 = (5, 15)$, $t^2 = (4, 10)$ (one can think of the players as the two companies A and B discussed in Section 1; the tasks then correspond to the two construction projects). Suppose that the players form two partial coalitions $\boldsymbol{r}^1$ and $\boldsymbol{r}^2$ of total weight 5 each so that player 1 contributes a unit of weight to $\boldsymbol{r}^1$ and 3 units of weight to $\boldsymbol{r}^2$, while player 2 contributes 4 units of weight to $\boldsymbol{r}^1$, and 2 units of weight to $\boldsymbol{r}^2$, that is, $CS = (\boldsymbol{r}^1, \boldsymbol{r}^2)$, where $\boldsymbol{r}^1 = (\frac{1}{4}, \frac{2}{3}), \boldsymbol{r}^2 = (\frac{3}{4}, \frac{1}{3})$. Both of these partial coalitions have weight 5, so each of them can successfully complete $t^1$, resulting in a payoff of 15 for each of them. Now, suppose that the players divide the gains using an imputation $\boldsymbol{x} = ((7, 8), (9, 6))$. Then, the total payoff obtained by player 2 is 14, so he can successfully deviate by withdrawing from both of these coalitions, and forming a single partial coalition of weight 5. This coalition can complete $t^1$ and receive a payoff of $15 > 14$. On the other hand, suppose that the players keep the same coalition structure, but distribute the gains as $\boldsymbol{y} = ((7, 8), (8, 7))$. Then player 2 can no longer gain by withdrawing from both of these coalitions. He is tempted to withdraw his resources from $\boldsymbol{r}^1$, as he can use these 4 units of weight to complete $t^2$ and earn $u^2 = 10 > 8$. However, if he does that, he can no longer get his share of payoffs from $\boldsymbol{r}^2$. Hence, in case of this deviation his total payoff will be $10 < 15$. Also, it is easy to see that player 2 cannot gain by deviating from $\boldsymbol{r}^2$ only, and player 1 is better off in $CS$ than he would be on his own. Hence, $(CS, \boldsymbol{y})$ is in the OCF-core of $\check{G}$.*

In some sense, Definition 7 takes a rather pessimistic, or *conservative*, view on what the members of the deviating group can expect to get from the non-deviators: indeed, in Example 2 as soon as player 2 withdraws from the partial coalition $\boldsymbol{r}^1 \in CS$ he expects to be thrown out of $\boldsymbol{r}^2$, even though $\boldsymbol{r}^2$ is not affected by this deviation. Therefore, in what follows, we will refer to the notion of profitable deviation introduced in Definition 7 as a *c-profitable deviation*, and to the corresponding notion of the core as the *conservative core*, or the *c-core*.

This definition is applicable when a deviation by an agent is interpreted by other agents as an indicator that this agent is not trustworthy, and therefore one should immediately stop all collaboration with him. While this kind of reaction is not unusual, there may be coalitions that are not affected by the deviation and may not want to punish the deviators. In this case, the deviators need to decide which of the existing coalitions to abandon and for which existing coalitions to keep their contribution intact. The members of these partial coalitions will react accordingly, sharing the payoff as before if they have not been affected by the deviation and punishing the deviators otherwise. Therefore, we refer to the corresponding notion of the core as *refined*. Before giving the formal definition, we first introduce a notion of agreement between two coalition structures.

**Definition 8.** *Given a set of agents $J \subseteq N$, we say that two coalition structures $CS$ and $CS'$ over $N$ agree outside of $J$ with respect to a function $f$ if $f$ is a bijection between the lists of partial coalitions $\{\boldsymbol{r}^i \in CS \mid \mathrm{supp}(\boldsymbol{r}^i) \nsubseteq J\}$ and $\{\boldsymbol{s}^\ell \in CS' \mid \mathrm{supp}(\boldsymbol{s}^\ell) \nsubseteq J\}$ such that $f(\boldsymbol{r}^i) = \boldsymbol{s}^\ell$ implies $r_j^i = s_j^\ell$ for all $j \notin J$. Further, we say that $CS$ and $CS'$ agree outside of $J$ if they agree outside of $J$ with respect to some function $f$.*

Intuitively, this definition says that if two coalition structures agree outside of $J$, then the contributions of all players *not* in $J$ to all partial coalitions must be the same under both outcomes. If $J$ is the set of deviators, this condition captures the fact that the deviation by the players in $J$ does not change the behavior of the non-deviators; the function $f$ is used to establish a correspondence between the partial coalitions involving the non-deviators before and after the deviation. For illustration, consider the following example.

**Example 3.** *Consider a game with three players $N = \{1, 2, 3\}$ and a coalition structure $CS = (\boldsymbol{q}^1, \boldsymbol{q}^2)$, where $\boldsymbol{q}^1 = (1, \frac{1}{2}, \frac{1}{2})$, $\boldsymbol{q}^2 = (0, \frac{1}{2}, \frac{1}{2})$. Let $CS' = (\boldsymbol{s}^1, \boldsymbol{s}^2, \boldsymbol{s}^3)$, where $\boldsymbol{s}^1 = (0, 0, \frac{1}{2})$, $\boldsymbol{s}^2 = (0, \frac{1}{2}, \frac{1}{2})$, $\boldsymbol{s}^3 = (1, \frac{1}{2}, 0)$. Intuitively, $CS'$ can be obtained from $CS$ when players 1 and 2 deviate by abandoning their joint project with player 3 and forming a coalition of their own. Set $J = \{1, 2\}$. It is not hard to see that $CS$ and $CS'$ agree outside of $J$ with respect to the function $f$ given by $f(\boldsymbol{q}^1) = \boldsymbol{s}^1$, $f(\boldsymbol{q}^2) = \boldsymbol{s}^2$. On the other hand, $CS$ and $CS'$ also agree outside of $J$ with respect to the function $f'$ given by $f'(\boldsymbol{q}^1) = \boldsymbol{s}^2$, $f'(\boldsymbol{q}^2) = \boldsymbol{s}^1$; this function assumes that when players 1 and 2 decided to deviate, player 1 withdrew his contribution to $\boldsymbol{q}^1$ and player 2 withdrew his contribution to $\boldsymbol{q}^2$.*

**Definition 9.** *Given an OCF-game $G = (N, v)$ and a set of agents $J \subseteq N$, let $(CS, \boldsymbol{x})$ and $(CS', \boldsymbol{y})$ be two outcomes such that $CS$ and $CS'$ agree outside of $J$ with respect to a function $f$. Suppose that for any partial coalition $\boldsymbol{s}^\ell \in CS'$ with $\operatorname{supp}(\boldsymbol{s}^\ell) \nsubseteq J$ and for all $j \in J$ we have $y_j^\ell = x_j^i$ if $\boldsymbol{r}^i = f^{-1}(\boldsymbol{s}^\ell)$ and $y_j^\ell = 0$ otherwise. Then we say that $(CS', \boldsymbol{y})$ is an r-profitable deviation of $J$ from $(CS, \boldsymbol{x})$ w.r.t. $f$ if for all $j \in J$ we have $p_j(CS', \boldsymbol{y}) > p_j(CS, \boldsymbol{x})$. Further, we say that $(CS', \boldsymbol{y})$ is an r-profitable deviation of $J$ from $(CS, \boldsymbol{x})$ if there exists a function $f$ such that $CS$ and $CS'$ agree outside of $J$ with respect to $f$ and $(CS', \boldsymbol{y})$ is an r-profitable deviation of $J$ from $(CS, \boldsymbol{x})$ w.r.t. $f$. We say that an outcome $(CS, \boldsymbol{x})$ is in the refined core, or the r-core, of $G$ if no subset of agents $J$ possesses an r-profitable deviation from it.*

In Definition 9, the bijection $f$ matches the partial coalitions in $CS$ and $CS'$ that involve non-deviators; the number of such coalitions is the same in both coalition structures. Moreover, the contribution of the non-deviators to the partial coalitions matched by $f$ is the same in $CS$ and $CS'$. Now, if also the deviators do not change their contribution to some partial coalition $\boldsymbol{r}$, they can claim their share of its payoff, as determined by $\boldsymbol{x}$. On the other hand, if the deviators change their contribution to $\boldsymbol{r}$, they are not entitled to any of its payoff. Observe that we allow the deviators to pick the "most favourable" bijection $f$ between $CS$ and $CS'$: for instance, in the context of Example 3 we would pick $f$ rather than $f'$, thereby allowing the deviators to claim their payoff from the coalition $(0, \frac{1}{2}, \frac{1}{2})$. In other words, we assume that the deviators will withdraw their contributions to disturb the non-deviators as little as possible.

**Example 4.** *Consider the game $\tilde{G}$ and the outcome $(CS, \boldsymbol{y})$ as described in Example 2. While it has been argued that player 2 cannot c-profitably deviate from $(CS, \boldsymbol{y})$, he can r-profitably deviate from it by withdrawing his weight from $\boldsymbol{r}^1$ and dedicating it to $t^2$. As he does not change his contribution to $\boldsymbol{r}^2$, he can still claim the payoff he gets from $\boldsymbol{r}^2$, so his total payoff is $10 + 7 = 17 > 15$.*

*On the other hand, suppose that players 1 and 2 both split their weights equally between two partial coalitions, forming the structure $CS' = (\boldsymbol{q}^1, \boldsymbol{q}^2)$, where $\boldsymbol{q}^1 = \boldsymbol{q}^2 = (\frac{1}{2}, \frac{1}{2})$. Clearly, both $\boldsymbol{q}^1$ and $\boldsymbol{q}^2$ have weight 5, so each of them can earn 15 by completing $t^1$. Now, suppose that the players distribute the gains using an imputation $\boldsymbol{x}' = ((3, 12), (12, 3))$. Now, both players earn 15, so none of them can benefit from withdrawing from both partial coalitions at the same time, and therefore the outcome $(CS', \boldsymbol{x}')$ is in the c-core. Moreover, if any of the players deviates from one coalition only, he does not have enough weight to complete any of the tasks, and therefore the outcome $(CS', \boldsymbol{x}')$ is also in the r-core.*

We now provide another example, which suggests that the set of profitable deviations allowed by Definition 9 may still be too small.

**Example 5.** *Consider again the game $\check{G}$ and a coalition structure $CS'' = (\boldsymbol{s}^1, \boldsymbol{s}^2)$, where player 1 contributes all of his weight to $\boldsymbol{s}^1$, while player 2 contributes 3 units of weight to $\boldsymbol{s}^1$ and 3 units of weight to $\boldsymbol{s}^2$, i.e., $\boldsymbol{s}^1 = (1, \frac{1}{2})$, $\boldsymbol{s}^2 = (0, \frac{1}{2})$. Observe that we have $v(\boldsymbol{s}^2) = 0$, as the total weight of $\boldsymbol{s}^2$ is 3 only. Now, consider an imputation $\boldsymbol{z} = ((3, 12), (0, 0))$. Note that player 2 could reduce his contribution to $\boldsymbol{s}^1$ by 2 units of weight without affecting the value of this coalition, and use this weight to boost the value of $\boldsymbol{s}^2$. However, this is not allowed by our definition of an r-profitable deviation, since as soon as player 2 alters his contribution to $\boldsymbol{s}^1$, he loses the payoff of 12 that he gets from $\boldsymbol{s}^1$. This does not mean, however, that the outcome $(CS'', \boldsymbol{z})$ is in the r-core of $\check{G}$: players 1 and 2 can collectively deviate to $((1, \frac{1}{6}), (0, \frac{5}{6}))$. If they share the payoff as $((4, 11), (0, 15))$, this will constitute an r-profitable deviation for both of them.*

Example 5 demonstrates that Definition 9, while being considerably more lax with respect to the deviators than Definition 7, can still be too strict: the deviators are punished as soon as they reduce their contribution to a coalition, irrespective of whether it affects the value of this coalition. In fact, according to Definition 9, the deviators would still be punished even if they *increase* their contribution to a partial coalition with non-deviators (though this type of deviation is, of course, unlikely). One way to fix this is to allow the deviators to claim their share of payoffs from a coalition $\boldsymbol{s}^\ell = f(\boldsymbol{r}^i)$ as long as $v(\boldsymbol{s}^\ell) = v(\boldsymbol{r}^i)$. However, the non-deviators can be even more generous to deviators. Indeed, it can be the case that after the deviators reduce their contribution to a particular partial coalition, this coalition is still able to perform some task, albeit of a smaller value. If the value of this task is still larger than the total amount of payoff originally received by the non-deviators from this partial coalition, the deviators could be allowed to claim the "leftover" payoff. In other words, this notion of deviation assumes that the non-deviators have no objection to switching tasks, and only care about the payoff they receive. While this may well be the case, it is quite optimistic of the deviators to expect this kind of reaction when they contemplate whether to deviate. Therefore, we refer to this notion of deviation as *o-profitable*, and call the corresponding solution concept the *optimistic core*, or the *o-core*.

**Definition 10.** *Given an OCF-game $G = (N, v)$ and a set of agents $J \subseteq N$, let $(CS, \boldsymbol{x})$ and $(CS', \boldsymbol{y})$ be two outcomes such that $CS$ and $CS'$ agree outside of $J$ with respect to a function $f$. Suppose also that for any partial coalition $\boldsymbol{s}^\ell \in CS'$ with $\mathrm{supp}(\boldsymbol{s}^\ell) \nsubseteq J$ we have $\sum_{j \in J} y_j^\ell = \max\{v(\boldsymbol{s}^\ell) - \sum_{k \in N \setminus J} x_k^i, 0\}$, where $\boldsymbol{r}^i = f^{-1}(\boldsymbol{s}^\ell)$. We say that $(CS', \boldsymbol{y})$ is an o-profitable deviation of $J$ from $(CS, \boldsymbol{x})$ w.r.t. $f$ if for all $j \in J$ we have $p_j(CS', \boldsymbol{y}) > p_j(CS, \boldsymbol{x})$. Further, we say that $(CS', \boldsymbol{y})$ is an* o-profitable deviation *of $J$ from $(CS, \boldsymbol{x})$ if there exists a function $f$ such that $CS$ and $CS'$ agree outside of $J$ with respect to $f$ and $(CS', \boldsymbol{y})$ is an* o-profitable deviation *of $J$ from $(CS, \boldsymbol{x})$ w.r.t. $f$. We say that an outcome $(CS, \boldsymbol{x})$ is in the* optimistic core, *or the* o-core, *of $G$ if no subset of agents $J$ has an o-profitable deviation from it.*

**Example 6.** *Consider again the game $\check{G}$ discussed in Examples 2, 4, and 5, and the outcome $(CS', \boldsymbol{x}')$, where $CS' = (\boldsymbol{q}^1, \boldsymbol{q}^2)$, $\boldsymbol{q}^1 = \boldsymbol{q}^2 = (\frac{1}{2}, \frac{1}{2})$, $\boldsymbol{x}' = ((3, 12), (12, 3))$, which was described in Example 4. Note that if player 2 reduces his contribution to $\boldsymbol{q}^1$ to 2, this coalition would still be able to earn 10 by focusing on task $t^2$. As player 1 only gets 3 units of payoff from $\boldsymbol{q}^1$ anyway, under our definition of an o-profitable deviation, player 2 is entitled to the remaining payoff from this modified partial coalition, i.e., $10 - 3 = 7$. He can then combine the unit of weight saved in* *this manner with the weight he contributes to $\boldsymbol{q}^2$, and embark on $t^2$ making a profit of* 10. *Thus, by abandoning $\boldsymbol{q}^2$ altogether and reducing his contribution to $\boldsymbol{q}^1$, player* 2 *can earn* $7 + 10 > 15$. *Thus, the outcome $(CS', \boldsymbol{x}')$ is not in the o-core of $\check{G}$.*

*In contrast, consider an outcome that combines $CS$ with a more symmetric payoff division scheme, such as, e.g., $\boldsymbol{y} = ((7, 8), (8, 7))$. Now, if player* 2 *reduces his contribution to $\boldsymbol{q}^1$ by* 1, *the resulting partial coalition can earn* 10 *by focusing on $t^2$. Of those payoffs, player* 1 *must receive* 7, *leaving* 3 *for player* 2. *While player* 2 *can still use his remaining weight to complete $t^2$, this will only give him a total profit of $10 + 3 = 13 < 15$, i.e., this deviation is not o-profitable. Similarly, we can show that withdrawing some of the resources from $\boldsymbol{q}^2$ and abandoning $\boldsymbol{q}^1$ is even less profitable for player* 2. *Finally, it is easy to see that player* 1 *does not have an o-profitable deviation either. Hence, the outcome $(CS', \boldsymbol{y})$ is in the o-core of $(\check{G})$.*

## 6. Core Characterization

In the previous section, we introduced three definitions of the core for overlapping coalition formation games. Among the three definitions of the core, the *c-core*, though in some sense conservative, is the closest to the traditional definition of the core in general NTU games (Osborne & Rubinstein, 1994). Indeed, unlike the other two definitions, it does not assume any interaction between the deviators and the non-deviators. This motivates us to study this overlapping core variant in more detail, which we proceed to do in this section and the next. To promote readability, in those two sections we will be referring to the *c-core* simply as "the core".

We start by providing a characterization of the set of outcomes in the core: essentially, an outcome is in the core if and only if under this outcome the total payments to each subset of agents match or exceed the maximum value that can be achieved by this subset. Our proof relies on some technical restrictions on the function $v$ that defines the game. In particular, we require $v$ to be continuous, monotone and bounded (observe that if a game is monotone and bounded, then $v^*(S) < \infty$ for any $S \subseteq N$), as well as to satisfy another natural restriction defined later. These assumptions allow us to avoid some pathological situations that may arise in our model at its generality, such as the supremum $v^*(N)$ being unachievable (e.g., if $v$ is strictly concave in one of its arguments, it can be the case that no finite coalition structure can achieve $v^*(N)$).

Specifically, we say that a game $(N, v)$ is *$U$-finite* if for any $(CS, \boldsymbol{x})$ such that $|CS| > U$ and $\boldsymbol{x} \in I(CS)$, there exists a $(CS', \boldsymbol{y})$ such that $|CS'| \leq U$, $\boldsymbol{y} \in I(CS')$, and $p_j(CS, \boldsymbol{x}) \leq p_j(CS', \boldsymbol{y})$ for all $j = 1, \ldots, n$ (i.e., for any outcome $(CS, \boldsymbol{x})$ with more than $U$ coalitions there exists another outcome $(CS', \boldsymbol{y})$ with at most $U$ coalitions that is weakly prefered to $(CS, \boldsymbol{x})$ by all agents). When this condition holds, we can assume that all coalition structures that arise in a game consist of at most $U$ partial coalitions. This is a natural restriction in many practical scenarios, as it might be difficult for agents to maintain a very complicated collaboration pattern. It holds when, for example, there is a bound on the number of partial coalitions each agent can be involved in. In general $U$-finiteness imposes some upper bound on the total number of partial coalitions with the same support that can occur. A natural example is provided by a class of games where for any two partial coalitions $\boldsymbol{r}, \boldsymbol{r}'$ such that $\mathrm{supp}(\boldsymbol{r}) = \mathrm{supp}(\boldsymbol{r}')$ and $r_j + r'_j \leq 1$ for any $j = 1, \ldots, n$, we have $v(\boldsymbol{r} + \boldsymbol{r}') \geq v(\boldsymbol{r}) + v(\boldsymbol{r}')$. Note that in such games we can assume that no coalition structure contains two partial coalitions with the same support $S$, as it is at least as profitable for the players in $S$ to merge these partial coalitions. (However, notice that this does not imply superadditivity, nor does it mean that the grand coalition necessarily emerges, as the criterion above refers only to coalitions with identical support.) Hence, any such game is $2^n$-finite.

**Remark 1.** *Note that in all of our results $U$ can be a function of $n$ (as long as $U(n) < \infty$). Alternatively, instead of imposing the condition of $U$-finiteness on $v(\cdot)$, we could restrict the set of allowed outcomes (or potential deviations) to coalition structures with at most $U$ partial coalitions. All of our results hold under this model as well.*

We now state and prove the first of our main results.

**Theorem 1.** *Given a game $(N, v)$, where $v$ is monotone, continuous, bounded, and $U$-finite for some $U \in \mathbb{N}$, an outcome $(CS, \boldsymbol{x})$ is in the c-core of $(N, v)$ if and only if for all $S \subseteq N$*

$$\sum_{j \in S} p_j(CS, \boldsymbol{x}) \geq v^*(S). \tag{1}$$

*Proof.* For the "if" direction, suppose that $(CS, \boldsymbol{x})$ satisfies $\sum_{j \in S} p_j(CS, \boldsymbol{x}) \geq v^*(S)$ for all $S \subseteq N$. Assume for the sake of contradiction that $(CS, \boldsymbol{x})$ is not in the core, i.e., there exists a set $S$, a coalition structure $CS_S \in \mathcal{CS}_S$ and an imputation $\boldsymbol{y} \in I(CS_S)$ such that $p_j(CS_S, \boldsymbol{y}) > p_j(CS, \boldsymbol{x})$ for all $j \in S$. Then we have $v(CS_S) = \sum_{j \in S} p_j(CS_S, \boldsymbol{y}) > \sum_{j \in S} p_j(CS, \boldsymbol{x}) \geq v^*(S)$, a contradiction with the way $v^*(S)$ was defined.

For the "only if" direction, consider an outcome $(CS, \boldsymbol{x})$ that does not satisfy (1); we will show that $(CS, \boldsymbol{x})$ is not in the core. To begin, set $\boldsymbol{p} = \boldsymbol{p}(CS, \boldsymbol{x})$, and assume $\sum_{j \in S} p_j < v^*(S)$ for some $S \subseteq N$. To show that $(CS, \boldsymbol{x})$ is not in the core, we will construct a set $S'$, a coalition structure $CS_{S'} \in \mathcal{CS}_{S'}$ and an imputation $\boldsymbol{y} \in I(CS_{S'})$ such that $p_j(CS_{S'}, \boldsymbol{y}) > p_j$ for all $j \in S'$. Fix a set $S$ that satisfies $\sum_{j \in S} p_j < v^*(S)$. Choose $\varepsilon$ small enough so that $\sum_{j \in S} p_j < v^*(S) - \varepsilon$, and let $\mathcal{CS}_S^\varepsilon = \{CS_S \in \mathcal{CS}_S \mid v(CS_S) \geq v^*(S) - \varepsilon\}$. By definition of $v^*(S)$, there is an infinite sequence of coalition structures $CS^{(t)}$ that satisfies $\lim_{t \to \infty} v(CS^{(t)}) = v^*(S)$, so the set $\mathcal{CS}_S^\varepsilon$ is non-empty. Given a coalition structure $CS_S \in \mathcal{CS}_S$, an imputation $\boldsymbol{y} \in I(CS_S)$ and a respective payoff vector $\boldsymbol{q} = \boldsymbol{p}(CS_S, \boldsymbol{y})$, define the *total loss* $TL(CS_S, \boldsymbol{q})$ of $(CS_S, \boldsymbol{q})$ as $\sum_{j:p_j > q_j}(p_j - q_j)$. Set $TL_{\min} = \inf\{TL(CS_S, \boldsymbol{q}) \mid CS_S \in \mathcal{CS}_S^\varepsilon, \boldsymbol{y} \in I(CS_S), \boldsymbol{q} = \boldsymbol{p}(CS_S, \boldsymbol{y})\}$. First, we prove that there exists a coalition structure $CS \in \mathcal{CS}_S^\varepsilon$ and an imputation $\boldsymbol{y} \in I(CS_S)$ that achieve the total loss of $TL_{\min}$.

**Lemma 1.** *Under the theorem's conditions, there exists a $CS_S \in \mathcal{CS}_S^\varepsilon$, an imputation $\boldsymbol{y} \in I(CS_S)$ and a payoff vector $\boldsymbol{q} = \boldsymbol{p}(CS_S, \boldsymbol{y})$ s.t. $TL(CS_S, \boldsymbol{q}) = TL_{\min}$.*

*Proof.* By definition of $TL_{\min}$, there exists an infinite sequence of coalition structures $CS_S^{(t)}$, $t = 1, \ldots, \infty$, and respective imputations $\boldsymbol{y}^{(t)}$, $t = 1, \ldots, \infty$, such that

$$\lim_{t \to \infty} TL(CS^{(t)}, \boldsymbol{p}(CS^{(t)}, \boldsymbol{y}^{(t)})) = TL_{\min}$$

and $CS_S^{(t)} \in \mathcal{CS}_S^\varepsilon$ for all $t = 1, \ldots, \infty$. As the game is $U$-finite, a coalition structure can be seen as a list of at most $U$ vectors in $[0, 1]^n$. By adding all-zero partial coalitions if necessary, we can assume that each coalition structure is a list of exactly $U$ vectors in $[0, 1]^n$, which are ordered lexicographically. As $v$ is monotone and bounded, there exists a $B > 0$ such that the value of each partial coalition in any of the $CS_S^{(t)}$ is between 0 and $B$. Consequently, each $\boldsymbol{y}^{(t)}$ corresponds to a vector in $[0, B]^{nU}$. Hence, the sequence $(CS_S^{(t)}, \boldsymbol{y}^{(t)})$, $t = 1, \ldots, \infty$ can be viewed as a subset of $[0, B]^K$ (for sufficiently large but finite value of $K$) and hence has a limit point, which we denote by $(CS^*, \boldsymbol{y}^*)$. It is easy to see that the limit of a sequence of coalition structures is a coalition structure, i.e., for any $\boldsymbol{r}^i \in CS^*$ we have $\boldsymbol{r} \in [0, 1]^n$, and for any $j = 1, \ldots, n$ it holds that $\sum_{i=1}^{U} r_j^i \leq 1$. Moreover, by continuity of $v$, the value of each partial coalition in $CS^*$ is the limit of the values of the respective partial coalitions in $CS_S^{(t)}$, $t = 1, \ldots, \infty$. From this, it is easy to see that $\boldsymbol{y}^*$ is in $I(CS^*)$. Also, as all $CS_S^{(t)}$ are in $\mathcal{CS}_S^c$, so is $CS^*$. Finally, as $p(\cdot, \cdot)$ and $TL(\cdot, \cdot)$ are continuous functions of their arguments, we conclude that $TL(CS^*, \boldsymbol{p}(CS^*, \boldsymbol{y}^*)) = TL_{\min}$. □

Continuing with the proof of our Theorem, let $(CS_S, \boldsymbol{y})$ be an outcome that satisfies $v(CS_S) \geq v^*(S) - \varepsilon$, $TL(CS_S, \boldsymbol{p}(CS_S, \boldsymbol{y})) = TL_{\min}$, whose existence is guaranteed by Lemma 1. Set $\boldsymbol{q} = \boldsymbol{p}(CS_S, \boldsymbol{y})$. Let us now construct a directed graph $\Gamma$ whose vertices are the agents and there is an edge from $j$ to $i$ if there exists a coalition in $CS_S$ containing both $j$ and $i$ such that under $\boldsymbol{y}$, agent $j$ gets a non-zero payoff from that coalition, i.e., for some $\boldsymbol{r}^k \in CS_S$ we have $r_j^k, r_i^k > 0$ and $y_j^k > 0$. Observe that if there is an edge $(j, i)$ in $\Gamma$, we can change $\boldsymbol{y}^k$ by increasing the payoff to $i$ by a small enough $\delta$ and decreasing the payoff to $j$ by the same value of $\delta$ without violating the constraints, i.e., we have $\boldsymbol{z} = (\boldsymbol{z}^1, \ldots, \boldsymbol{z}^t) \in I(CS_S)$, where $\boldsymbol{z}^l = \boldsymbol{y}^l$ for $l \neq k$ and $\boldsymbol{z}^k = (y_1^k, \ldots, y_j^k - \delta, \ldots, y_i^k + \delta, \ldots, y_n^k)$. Now, color all vertices of $\Gamma$ as follows: a vertex $j$ is red if the agent $j$ is underpaid under $\boldsymbol{y}$, i.e., $q_j < p_j$, white if $j$ is indifferent, i.e., $q_j = p_j$, and green if he is overpaid, i.e., $q_j > p_j$. As $\sum_{j \in S} p_j < v^*(S) - \varepsilon$ and $\sum_{j \in S} q_j = v(CS_S) \geq v^*(S) - \varepsilon$, the graph contains at least one green vertex. As argued above, if there is a path from a green vertex $j$ to a red vertex $i$, we can transfer a small amount of payoff from $j$ to $i$ and hence decrease the total loss, which is a contradiction with our choice of $(CS_S, \boldsymbol{y})$. Hence, given an arbitrary green vertex $j$, the set of all vertices reachable from $j$ in the graph, which we denote by $R(j)$, can only contain green or white vertices.

We would now like to argue that the agents in $R(j)$ can successfully deviate from $(CS, \boldsymbol{x})$. Indeed, let $CS'$ be the coalition structure that consists of the coalitions that the agents in $R(j)$ form among themselves in $CS_S$. Clearly, the value of $CS'$ is equal to the total value of the coalitions formed by these agents in $CS_S$. Note also that under $(CS_S, \boldsymbol{y})$, the agents in $R(j)$ do not get any payoffs from coalitions that involve agents not in $R(j)$. Indeed, suppose that an $i \in R(j)$ gets a non-zero payoff from a coalition that involves an agent $k \notin R(j)$. Then in $\Gamma$ there is an edge from $i$ to $k$, a contradiction with how $R(j)$ was constructed. In other words, in $CS_S$, the payoffs that the agents in $R(j)$ get come only from the coalitions that they form among themselves, and yet these agents are all green or white, i.e., each of them is doing no worse than what he was doing under $CS$, and some of them (in particular, agent $j$) are doing strictly better. To finish the proof, let the agents in $R(j)$ distribute the payoffs in the same way as in $(CS_S, \boldsymbol{y})$, except that player $j$ transfers a small fraction of his payoffs to each of the white players in $R(j)$ (this is possible by construction). The last step ensures that each agent in $R(j)$ is strictly better off than in $(CS, \boldsymbol{x})$. This demonstrates that $(CS, \boldsymbol{x})$ is not in the core, as required. □

**Remark 2.** *Note that we did not have to make use of the additional restrictions we imposed on $v$ to prove the "if" direction of the theorem (these are used in the proof of Lemma 1). Hence, this implication holds for an arbitrary $G$.*

It is easily verifiable that Theorem 1 holds in the non-overlapping case with coalition structures as well. The result is trivial to prove in that setting, as each agent's payoffs come from just one coalition; in contrast, we had to use more involved combinatorial arguments for transferring payoffs among agents. We also get the following interesting result as a corollary:

**Corollary 1.** *By setting $S = N$ in the statement of Theorem 1, we conclude that any outcome in the* c-core *maximizes the social welfare.*

We now turn our attention to characterizing the set of coalition structures $CS$ that admit payoff allocations $\boldsymbol{x}$ such that the corresponding tuple $(CS, \boldsymbol{x})$ belongs to the core. That is, while in Theorem 1 we saw a necessary and sufficient condition for a tuple $(CS, \boldsymbol{x})$ to belong to the core, suppose that we are now only given a structure $CS = (\boldsymbol{r}^1, \ldots, \boldsymbol{r}^k)$ and we want to check whether there exists *some* payoff allocation $\boldsymbol{x}$ such that $(CS, \boldsymbol{x})$ belongs to the core. Our characterization can be seen as a generalization of the notion of *balancedness* in the context of overlapping coalition formation. In the classic setting, the analogous question is "when does the grand coalition admit a payoff allocation in the core", answered by Bondareva (1963) and Shapley (1967). Before we proceed to our result, we define balancedness with respect to a coalition structure.

**Definition 11.** *Fix a coalition structure $CS = (\boldsymbol{r}^1, \ldots, \boldsymbol{r}^k)$, $k \in \mathbb{N}$, and let $K = \{1, ..., k\}$. A collection of numbers $\{\lambda_S\}_{S \subseteq N}, \{\mu_i\}_{i \in K}$ is called* balanced w.r.t. the given coalition structure $CS$ *if and only if $\lambda_S \geq 0$ for all $S$, and $\sum_{S: j \in S} \lambda_S + \mu_i = 1$ for all $i \in K, j \in \mathrm{supp}(\boldsymbol{r}^i)$.*

**Definition 12.** *A game is called* balanced w.r.t. a coalition structure $CS = (\boldsymbol{r}^1, ..., \boldsymbol{r}^k)$ *if and only if for every collection $\{\lambda_S\}_{S \subseteq N}, \{\mu_i\}_{i \in K}$ that is balanced w.r.t. $CS$ it holds that $\sum_S \lambda_S v^*(S) + \sum_{i=1}^k \mu_i v(\boldsymbol{r}^i) \leq v^*(N)$.*

The proof of the following theorem is based on LP-duality, and relies on the characterization result of Theorem 1; furthermore, the proof illustrates that the condition of balancedness introduced above arises rather naturally.

**Theorem 2.** *Let $(N, v)$ be an OCF-game, where $v$ is monotone, continuous, bounded, and $U$-finite for some $U \in \mathbb{N}$ and consider a coalition structure $CS = (\boldsymbol{r}^1, ..., \boldsymbol{r}^k)$, for some $k \in \mathbb{N}$. There exists an imputation $\boldsymbol{x}$ s.t. $(CS, \boldsymbol{x})$ belongs to the* c-core *if and only if the game is balanced w.r.t. $CS$.*

*Proof.* Suppose there exists a payoff allocation $\boldsymbol{x}$ such that $(CS, \boldsymbol{x})$ belongs to the core, and let $K = \{1, \ldots, k\}$. Then the following linear program (denoted as LP) has an optimal solution:

$$
\begin{array}{lll}
\min & \sum_{i \in K, j \in N} x_{ij} & \\
\text{s.t.} & \sum_{j \in S} \sum_{i: j \in \mathrm{supp}(\boldsymbol{r}^i)} x_{ij} \geq v^*(S) & \forall S \subseteq N \\
& \sum_j x_{ij} = v(\boldsymbol{r}^i) & \forall i \in K
\end{array}
$$

The first constraint expresses the condition of Theorem 1, and the second the fact that the payoff of each partial coalition needs to be distributed exactly. Note that we have no variables $x_{ij}$ if $j \notin \mathrm{supp}(\boldsymbol{r}^i)$—recall Definition 4. These are precisely the conditions that need to be satisfied for $(CS, \boldsymbol{x})$ to be in the core and clearly the optimal value of the LP is $v^*(N)$ (using the first constraint and Corollary 1). By the LP-duality theorem, this means that the dual program also has an optimal solution of value $v^*(N)$. The dual is given by:

$$
\begin{array}{lll}
\max & \sum_S \lambda_S v^*(S) + \sum_{i=1}^k \mu_i v(\boldsymbol{r}^i) & \\
\text{s.t.} & \sum_{S: j \in S} \lambda_S + \mu_i = 1 & \forall i \in K, j \in \mathrm{supp}(\boldsymbol{r}^i) \\
& \lambda_S \geq 0 & \forall S \subseteq N
\end{array}
$$

Hence for any feasible solution of the dual, the value of the objective function is at most $v^*(N)$, which implies that for any balanced collection $\{\lambda_S\}_{S \subseteq N}, \{\mu_i\}_{i \in K}$, it holds that $\sum_S \lambda_S v^*(S) + \sum_{i=1}^k \mu_i v(\boldsymbol{r}^i) \leq v^*(N)$.

For the other direction, suppose that for any balanced collection, the above holds. This means that for any feasible solution, the value of the dual is at most $v^*(N)$. Therefore the dual is both bounded and feasible (setting $\mu_i = 1$ and the rest to 0 is feasible), which implies that it has an optimal solution. But then the primal program also has an optimal solution $\boldsymbol{x}$ and this means by Theorem 1 that $(CS, \boldsymbol{x})$ belongs to the core. $\qquad\square$

**Remark 3.** *In the traditional superadditive setting, the condition of balancedness is somewhat simpler and more intuitive. In our setting, the characterization leads to a slightly more complicated expression, essentially due to the fact that the linear program that describes core allocations for each coalition structure requires a larger set of constraints.*

## 7. Convex OCF-Games Have a Non-Empty Core

In this section, we first generalize the notion of convexity to OCF-games and then proceed to show that it provides a sufficient condition for non-emptiness of the c-core.

Recall that for classical TU-games convexity means that for $R \subseteq N$ and $S \subseteq T \subseteq N \setminus R$ it holds that $v(S \cup R) - v(S) \leq v(T \cup R) - v(T)$. Thus, convexity in the classic TU-games setting means that it is more useful for a coalition $R$ to join a larger coalition than a smaller one. We now apply this intuition to our setting (recall that $\mathcal{F}(S)$ denotes the set of all feasible agreements for $S$):

**Definition 13.** *An OCF-game $G = (N, v)$ is convex if for each $R \subseteq N$ and $S \subseteq T \subseteq N \setminus R$ the following condition holds: for any $(CS^S, \boldsymbol{x}^S) \in \mathcal{F}(S)$, any $(CS^T, \boldsymbol{x}^T) \in \mathcal{F}(T)$, and any $(CS^{S \cup R}, \boldsymbol{x}^{S \cup R}) \in \mathcal{F}(S \cup R)$ that satisfies $p_j(CS^{S \cup R}, \boldsymbol{x}^{S \cup R}) \geq p_j(CS^S, \boldsymbol{x}^S) \; \forall j \in S$, there exists an outcome $(CS^{T \cup R}, \boldsymbol{x}^{T \cup R}) \in \mathcal{F}(T \cup R)$ s.t.*

$$p_j(CS^{T \cup R}, \boldsymbol{x}^{T \cup R}) \geq p_j(CS^T, \boldsymbol{x}^T) \quad \forall j \in T, \text{ and}$$
$$p_j(CS^{T \cup R}, \boldsymbol{x}^{T \cup R}) \geq p_j(CS^{S \cup R}, \boldsymbol{x}^{S \cup R}) \quad \forall j \in R.$$

This definition is similar in flavour to that provided by Suijs and Borm (1999), where a generalization of convexity is defined in the context of stochastic cooperative games. The intuition behind this definition is as follows: Consider two fixed agreements, one on $S$ and one on $T$ respectively. Any time that there is a feasible agreement on $S \cup R$ that the members of $S$ do not object to compared to their own agreement (i.e., all members of $S$ are weakly better off than in their previous agreement), then there is a feasible agreement on $T \cup R$ such that (i) the members of $T$ do not object to this agreement, compared to the previous agreement on $T$ and (ii) the members of $R$ weakly prefer this agreement to the agreement on $S \cup R$.

We note that a different notion of convexity has been defined for fuzzy games by Branzei, Dimitrov, & Tijs (2003). That definition deals with the marginal contribution of a partial coalition when joining another existing partial coalition, where the result of the join is a new partial coalition. We, on the other hand, quantify the marginal contribution of adding a set of players $R$, to a set of players $T$, w.r.t. the best overlapping coalition structure that the set $R \cup T$ can form. Secondly, the definition of Branzei et al., as well as the classic definition of convexity, simply enforce a property on the function $v(\cdot)$, concerning the marginal contribution $v(R \cup T) - v(T)$. In our case, our games are not fully transferable and hence we cannot simply talk about the difference in values. Instead, our definition has to enforce the existence of a coalition structure on $R \cup T$ such that individually every player is at least as well-off as in the coalition structure over $R \cup S$, where $S \subseteq T$.

We now show that convexity is a sufficient condition for the non-emptiness of the core, in analogy to the classic result on convex TU-games (Shapley, 1971).

**Theorem 3.** *If an OCF-game $G = (N, v)$ is convex and $v$ is continuous, bounded, monotone and $U$-finite for some $U \in \mathbb{N}$, then the c-core of this game is not empty.*

*Proof.* Let $G = (N, v)$ be a convex OCF-game. For any $S \subseteq N$, let $G^S$ be the restriction of $G$ on $S$. To prove the theorem, we explicitly construct an outcome $(CS, \boldsymbol{x})$, $\boldsymbol{x} \in I(CS)$, and show that it belongs to the core of $G$: Fix an arbitrary ordering of the players $1, 2, \ldots, n-1, n$. The construction takes place in rounds. First, let $\hat{p}_1 = v^*(\{1\})$, $\hat{p}_2 = v^*(\{2\})$; by assumptions of the theorem and using arguments similar to those in the proof of Lemma 1, there exist coalition structures in $\mathcal{CS}_{\{1\}}, \mathcal{CS}_{\{2\}}$ that achieve these payoffs. Let $CS^1$ be the structure that achieves this for player 1 in $G^{\{1\}}$, and let $\boldsymbol{x}^1$ be the corresponding imputation. We know that there exists at least one coalition structure $CS^2 \in \mathcal{CS}_{\{1,2\}}$ and a corresponding imputation $\boldsymbol{x}^2$ such that $p_1(CS^2, \boldsymbol{x}^2) \geq \hat{p}_1$, $p_2(CS^2, \boldsymbol{x}^2) \geq \hat{p}_2$ (e.g., take the union of payoff-maximizing structures in $G^{\{1\}}$ and $G^{\{2\}}$, and combine the corresponding imputations). If there exist more than one such feasible agreement, we pick the one most preferred by player 2. More formally, we choose a feasible agreement $(CS^2, \boldsymbol{x}^2)$ that maximizes the payoff $p_2(CS^2, \boldsymbol{x}^2)$ (which will be at least $\hat{p}_2$) over all feasible agreements on $\{1, 2\}$ subject to $p_1(CS^2, \boldsymbol{x}^2) \geq p_1(CS^1, \boldsymbol{x}^1)$ (by our assumptions on $v(\cdot)$, this maximum exists).

Now, let $\hat{p}_3$ be the maximum payoff that agent 3 can get in $G^{\{3\}}$. Again, there exists at least one coalition structure $CS^3$ in $\mathcal{CS}_{\{1,2,3\}}$ and a corresponding imputation $\boldsymbol{x}^3$ such that agents 1, 2 are (weakly) better off than in $(CS^2, \boldsymbol{x}^2)$, and 3 is also weakly better off than being on its own. If there exist more than one such feasible agreement, we pick one that maximizes 3's payoff, i.e., we pick an agreement $(CS^3, \boldsymbol{x}^3)$ so that $p_3(CS^3, \boldsymbol{x}^3)$ is maximized over all agreements on $\{1, 2, 3\}$ subject to the constraints $p_1(CS^3, \boldsymbol{x}^3) \geq p_1(CS^2, \boldsymbol{x}^2)$, $p_2(CS^3, \boldsymbol{x}^3) \geq p_2(CS^2, \boldsymbol{x}^2)$.

Continuing in the same manner, at every round $k$ we pick an outcome $(CS^k, \boldsymbol{x}^k)$ that maximizes $p_k(CS^k, \boldsymbol{x}^k)$ subject to constraints $p_i(CS^k, \boldsymbol{x}^k) \geq p_i(CS^{k-1}, \boldsymbol{x}^{k-1})$ for $i \in \{1, ..., k-1\}$; the assumptions on $v(\cdot)$ ensure that all these maxima exist. In the end, we obtain a feasible agreement $(CS^n, \boldsymbol{x}^n)$ on $N$ in which all the agents are weakly better off than on their own, as well as weakly better off compared to the agreements of the previous rounds.

We now show that $(CS^n, \boldsymbol{x}^n)$ belongs to the core of $G$. For this it suffices to prove the following stronger claim.

**Claim 1.** *For $k = 1, \ldots, n$, the feasible agreement $(CS^k, \boldsymbol{x}^k)$ belongs to the core of the game $G^{\{1,\ldots,k\}}$.*

*Proof.* We prove this by induction. For $k = 1$, it is obvious that $(CS^1, \boldsymbol{x}^1)$ belongs to the core of $G^{\{1\}}$.

Now, suppose that for some $m$, $2 \leq m \leq n$, we have $(CS^k, \boldsymbol{x}^k) \in \text{core}(G^{\{1,\ldots,k\}})$ for all $k < m$. We will prove that $(CS^m, \boldsymbol{x}^m)$ is in the core of $G^{\{1,\ldots,m\}}$.

Suppose, for the sake of contradiction, that this is not the case. Then there is a subset $S \subseteq \{1, ..., m\}$ and $(CS^*, \boldsymbol{x}^*) \in \mathcal{F}(S)$ such that

$$p_i(CS^*, \boldsymbol{x}^*) > p_i(CS^m, \boldsymbol{x}^m) \ \forall i \in S. \tag{2}$$

We consider three different cases for the members of $S$:

**Case 1:** $m \notin S$. In this case we know by construction that for all $i \in \{1, \ldots, m-1\}$ we have $p_i(CS^m, \boldsymbol{x}^m) \geq p_i(CS^{m-1}, \boldsymbol{x}^{m-1})$, which implies that $p_i(CS^*, \boldsymbol{x}^*) > p_i(CS^{m-1}, \boldsymbol{x}^{m-1})$ for all $i \in S$. Hence, the tuple $(CS^*, \boldsymbol{x}^*)$ is a deviation that makes the members of $S$ strictly better off than in the agreement $(CS^{m-1}, \boldsymbol{x}^{m-1})$. But this is a contradiction since by induction $(CS^{m-1}, \boldsymbol{x}^{m-1}) \in$ core$(G^{\{1,\ldots,m-1\}})$.

**Case 2:** $S = \{1, \ldots, m\}$. Now we will get a contradiction with how we constructed $(CS^m, \boldsymbol{x}^m)$. Indeed, we chose $(CS^m, \boldsymbol{x}^m)$ to maximize $p_m(CS^m, \boldsymbol{x}^m)$ subject to the constraints $p_i(CS^m, \boldsymbol{x}^m) \geq p_i(CS^{m-1}, \boldsymbol{x}^{m-1})$ for all $i = 1, \ldots, m-1$. However, by (2), the outcome $(CS^*, \boldsymbol{x}^*)$ also satisfies these constraints and provides a higher payoff to $m$ than $(CS^m, \boldsymbol{x}^m)$ does, a contradiction.

**Case 3:** $S = S' \cup \{m\}$, where $S'$ is a strict subset of $\{1, \ldots, m-1\}$. In this case we will utilize convexity. Let $CS'$ be the coalition structure that consists of the singleton coalitions for all agents of $S'$, and let $\boldsymbol{x}'$ be the corresponding imputation. By construction, $(CS^*, \boldsymbol{x}^*)$ is a feasible agreement on $S' \cup \{m\}$ such that $p_i(CS^*, \boldsymbol{x}^*) \geq p_i(CS', \boldsymbol{x}')$ for all $i \in S'$. Let $T = \{1, \ldots, m-1\}$. Since $(CS^{m-1}, \boldsymbol{x}^{m-1}) \in \mathcal{F}(T)$, by applying Def. 13 for $S' \subseteq T$ and with $R = \{m\}$, we get that there exists a feasible agreement $(CS, \boldsymbol{x})$ on $T \cup \{m\} = \{1, \ldots, m\}$ such that $p_i(CS, \boldsymbol{x}) \geq p_i(CS^{m-1}, \boldsymbol{x}^{m-1})$ for $i = 1, \ldots, m-1$, and $p_m(CS, \boldsymbol{x}) \geq p_m(CS^*, \boldsymbol{x}^*)$. But then by (2) above we get that $p_m(CS, \boldsymbol{x}) > p_m(CS^m, \boldsymbol{x}^m)$, a contradiction with how we chose $(CS^m, \boldsymbol{x}^m)$. □

Applying Claim 1 with $k = n$, we get that the core of $G$ is non-empty. □

In the traditional setting, if a game is represented using oracle access for $v(S)$, there is a trivial algorithm for computing an element of the core in convex games. Indeed, one can set the payoff vector to be the vector of the marginal contributions of the agents for an arbitrary permutation of the set of agents. In our setting, our proof does yield a procedure for constructing an element of the core, though not a polynomial-time one. Our procedure requires solving a series of optimization questions, which for arbitrary convex games are NP-hard. In the future, we would like to find classes of convex games where our proof yields a polynomial-time algorithm. In particular, looking at our proof, this would be true for games in which we can solve in polynomial time the following problem: Given a set of agents $S \subseteq N$, a feasible agreement on $S$, an outcome $(CS, \boldsymbol{x})$, and an agent $k \notin S$, find a feasible agreement $(CS', \boldsymbol{y})$ on $S \cup \{k\}$ that maximizes $p_k(CS', \boldsymbol{y})$ subject to the constraints $p_j(CS', \boldsymbol{y}) \geq p_j(CS, \boldsymbol{x})$.

# 8. Properties of the Three Cores

Following the detailed study of the *c-core* stability concept in the previous two sections, in this section we further explore the properties of our three notions of the OCF-core. In particular, we investigate the relationships among these notions, and study the effects of allowing overlapping coalition formation on the stability of the underlying game. We also compare our OCF model and notions of the core to the fuzzy games setting and the notion of the fuzzy core (Aubin, 1981).

We start by exploring the connection between stability and social welfare maximization in TTGs. As demonstrated earlier in the paper, in OCF-games these two properties are closely related. Indeed, Theorem 1 and Corollary 1 show that any outcome in the c-core of an OCF-game maximizes the social welfare as long as the characteristic function of the game satisfies a number of technical conditions; by Theorem 5 below the same holds for the r-core and the o-core. However, as one of these conditions is continuity, this result does not directly apply to TTGs. While the proof of Theorem 1 can be adapted to work for the TTG setting, there also exists a direct proof for the following theorem.

**Theorem 4.** *For any TTG $G = (N; \boldsymbol{w}; \boldsymbol{t})$ and any outcome $(CS, \boldsymbol{x}) \in$ c-core$(\check{G})$, we have $v(CS) \geq v(CS')$ for any coalition structure $CS' \in \mathcal{CS}_N$.*

*Proof.* Fix an outcome $(CS, \boldsymbol{x}) \in$ c-core$(\check{G})$, and let $\boldsymbol{p}$ be the payoff vector that corresponds to $(CS, \boldsymbol{x})$. Suppose that there exists a coalition structure $CS' \in \mathcal{CS}_N$ such that $v(CS') > v(CS)$. Let $CS' = (\boldsymbol{r}^1, \dots, \boldsymbol{r}^k)$. For $j = 1, \dots, k$, let $z^j$ be the total weight of the partial coalition $\boldsymbol{r}^j$, i.e., set $z^j = r_1^j w_1 + \dots + r_n^j w_n$.

Now, consider a coalition structure $CS'' = (\boldsymbol{q}^1, \dots, \boldsymbol{q}^k)$ given by $q_i^j = z^j/w(N)$ for all $i \in N$, all $j = 1, \dots, k$; note that we have $\sum_{j=1}^k q_i^j \leq 1$. The total weight of a partial coalition $\boldsymbol{q}^j$ can be computed as $\sum_{i \in N} q_i^j w_i = z^j$. Therefore, $\boldsymbol{q}^j \in CS''$ can accomplish the same task as $\boldsymbol{r}^j \in CS'$, and hence $v(CS'') = v(CS') > v(CS)$. Now, observe that since in $CS''$ all players contribute to all partial coalitions, there are no restrictions on how the value of $CS''$ can be distributed among the players. In particular, we can set $\delta = \frac{v(CS'')-v(CS)}{n}$, and construct an imputation $\boldsymbol{y} \in I(CS'')$ by setting $y_i^j = \frac{v(\boldsymbol{r}^j)}{v(CS')}(p_i + \delta)$. Indeed, we have $\sum_{i \in N} y_i^j = v(\boldsymbol{r}^j)$, $\sum_{j=1}^k y_i^j = p_i + \delta$. Now, it is clear that the entire set of agents $N$ can deviate from $(CS, \boldsymbol{x})$ to $(CS'', \boldsymbol{y})$; as they all deviate simultaneously, this is a c-profitable deviation, a contradiction with $(CS, \boldsymbol{x})$ being in the c-core of $\check{G}$. ☐

The discussion in Section 5.2 suggests a natural relationship between the three notions of a successful deviation, and, consequently, between the three cores. (In what follows, we refer to the outcomes in the c-core, r-core and o-core as *c-stable*, *r-stable* and *o-stable*, respectively.)

**Theorem 5.** *For any OCF-game $G$, we have o-core$(G) \subseteq$ r-core$(G) \subseteq$ c-core$(G)$. Moreover, these containments can be strict, i.e., there exists an OCF-game $G$ such that o-core$(G) \subset$ r-core$(G) \subset$ c-core$(G)$.*

*Proof.* Observe that any c-profitable deviation can be viewed as an r-profitable deviation in which all players abandon all coalitions they contributed to. Similarly, any r-profitable deviation corresponds to an o-profitable deviation where whenever a deviator changes his contribution to coalition, he withdraws all of his resources from it; note that, as illustrated by Example 5, the deviators' payoff in this o-profitable deviation can be strictly higher than in the original r-profitable deviation. It follows that any outcome that is r-stable is also c-stable, and any outcome that is o-stable is also r-stable, thus proving the first part of the theorem.

To prove the second part of the theorem, consider the game $\check{G}$ described in Examples 2, 4, 5 and 6. We have demonstrated that the outcome $(CS, \boldsymbol{x})$ is in c-core$(\check{G}) \setminus$ r-core$(\check{G})$ and that the outcome $(CS', \boldsymbol{x}')$ is in r-core$(\check{G}) \setminus$ o-core$(\check{G})$. ☐

Theorem 5 shows that our three notions of stability can be substantially different with respect to *individual outcomes*. However, it does not exclude the possibility that they are equivalent when seen as notions of stability of the *entire game*, i.e., that for any OCF-game $G$ we have c-core$(G) \neq \emptyset$ iff r-core$(G) \neq \emptyset$ iff o-core$(G) \neq \emptyset$. We will now show that this is not the case. The games used in the proofs of the following two propositions are not threshold task games. However, they, too, can be described in terms of agents' weights and tasks.

**Proposition 1.** *There exists an OCF-game $G$ such that* c-core$(G) \neq \emptyset$ *while* r-core$(G) = \emptyset$.

*Proof.* Consider an OCF-game $G = (N, v)$ with seven agents $N = \{1, \ldots, 7\}$ whose weights are given by $\boldsymbol{w} = (1, 1, 1, 1, 3, 3, 3)$, and two task types $t^1$ and $t^2$ with values 100 and 2, respectively. The first task can be completed in any of the following four ways:

- 1 unit of player 1's weight and 2 units of player 5's weight;

- 1 unit of player 2's weight and 2 units of player 6's weight;

- 1 unit of player 3's weight and 2 units of player 7's weight;

- 1 unit of player 4's weight and 2 units of weight from either of the players 5, 6, or 7.

That is, $v(\boldsymbol{r}) = 100$ if $w_i r_i \geq 1$ and $w_j r_j \geq 2$, where

$$(i, j) \in \{(1, 5), (2, 6), (3, 7), (4, 5), (4, 6), (4, 7)\}.$$

The second task $t^2$ requires 2 units of weight in total from players 5, 6 and 7.

Consider a coalition structure $CS = (\boldsymbol{r}^1, \boldsymbol{r}^2, \boldsymbol{r}^3, \boldsymbol{r}^4)$, given by

$$\boldsymbol{r}^1 = (1, 0, 0, 0, \frac{2}{3}, 0, 0), \quad \boldsymbol{r}^2 = (0, 1, 0, 0, 0, \frac{2}{3}, 0),$$
$$\boldsymbol{r}^3 = (0, 0, 1, 0, 0, 0, \frac{2}{3}), \quad \boldsymbol{r}^4 = (0, 0, 0, 0, \frac{1}{3}, \frac{1}{3}, 0).$$

That is, partial coalitions $\boldsymbol{r}^1$, $\boldsymbol{r}^2$ and $\boldsymbol{r}^3$ successfully complete $t^1$, while $\boldsymbol{r}^4$ successfully completes $t^2$. Consider also an imputation $\boldsymbol{x} \in \mathcal{I}(CS)$ given by

$$\boldsymbol{x}^1 = (0, 0, 0, 0, 100, 0, 0), \quad \boldsymbol{x}^2 = (0, 0, 0, 0, 0, 100, 0),$$
$$\boldsymbol{x}^3 = (0, 0, 0, 0, 0, 0, 100), \quad \boldsymbol{x}^4 = (0, 0, 0, 0, 1, 1, 0).$$

Let $\boldsymbol{p}$ be the payoff vector that corresponds to $\boldsymbol{x}$: we have $p_1 = p_2 = p_3 = p_4 = 0$, $p_5 = p_6 = 101$, $p_7 = 100$. It is not hard to see that $(CS, \boldsymbol{x}) \in$ c-core$(G)$. Indeed, suppose for the sake of contradiction that there is a set of players $J$ that can c-profitably deviate from $(CS, \boldsymbol{x})$. Since $(CS, \boldsymbol{x})$ maximizes the social welfare, the deviation cannot be simultaneously profitable for all players in $N$, so $|J| < 7$. Moreover, $J$ cannot contain 2 or more players from the set $S = \{5, 6, 7\}$: indeed, if one of these players deviates, he loses 100 units of payoff, which can only be replaced if he forms a coalition with 4. However, since 4 cannot form two distinct coalitions of value 100 each, this is not possible. Therefore, $J$ cannot contain any of the players in the set $S$: each of these players already gets the maximum payoff from $t^1$, and, since the other two players from $S$ are not in $J$, the set of deviators does not have enough resources for $t^2$. Finally, there is no c-profitable deviation for players in $N \setminus S$, as no task can be completed by agents in $N \setminus S$ only.

We will now show that the r-core of $G$ is empty. Suppose otherwise, and let $(CS', \boldsymbol{y})$ be an outcome in the r-core of $G$. Let $\boldsymbol{p}$ be the payoff vector that corresponds to $\boldsymbol{y}$. It is not hard to show that any outcome in the r-core of $G$ maximizes the social welfare; the proof is similar to that of Theorem 4. Hence, we can assume without loss of generality that $CS = (\boldsymbol{q}^1, \boldsymbol{q}^2, \boldsymbol{q}^3, \boldsymbol{q}^4)$ with $v(\boldsymbol{q}^1) = v(\boldsymbol{q}^2) = v(\boldsymbol{q}^3) = 100$ and $v(\boldsymbol{q}^4) = 2$, and, moreover, $q_5^1 \geq \frac{2}{3}, q_6^2 \geq \frac{2}{3}, q_7^3 \geq \frac{2}{3}$. It follows that either (a) $q_1^1 = q_2^2 = q_3^3 = 1$ or (b) $q_4^j = 1$ for some $j \in \{1, 2, 3\}$ and $q_i^i = 1$ for $i \in \{1, 2, 3\}$, $i \neq j$. We say that a player $i$ is *useful* for a coalition $\boldsymbol{r}$ if $v(\boldsymbol{r}') < v(\boldsymbol{r})$, where $\boldsymbol{r}'$ is given by $r_i' = 0$, $r_j' = r_j$ for all $j \neq i$. Observe that in an r-stable outcome no player can get any payoff from a partial coalition for which he is not useful: otherwise the other members of that coalition, who can complete the corresponding task on their own, can r-profitably deviate. We will now show that we have $p_1 = \ldots = p_4 = 0$ both in case (a) and in case (b). Observe that by the argument above player 1 can get payoff from $\boldsymbol{q}^1$ only, player 2 can get payoff from $\boldsymbol{q}^2$ only, player 3 can get payoff from $\boldsymbol{q}^3$ only, and player 4 can get payoff from exactly one of the coalitions $\boldsymbol{q}^1$, $\boldsymbol{q}^2$, and $\boldsymbol{q}^3$.

In case (a), we clearly have $p_4 = 0$, as player 4 is not useful for any coalition in $CS'$. Now, if, e.g., $y_1^1 > 0$, then $y_5^1 < 100$, and players 4 and 5 can r-profitably deviate by forming a coalition that performs $t^1$. Hence $y_1^1 = y_2^2 = y_3^3 = 0$, and therefore $p_1 = p_2 = p_3 = 0$. In case (b), assume without loss of generality that $q_4^1 = 1$. Then $p_1 = 0$, as player 1 is not useful for any coalition in $CS'$, so $y_4^1 = 0$, since otherwise players 1 and 5 can r-profitably deviate, and, consequently, $p_4 = 0$. This implies that also $y_2^2 = y_3^3 = 0$: if, e.g., $y_2^2 > 0$, then $y_6^2 < 100$, and players 4 and 6 can r-profitably deviate by forming a coalition that performs $t^1$. Hence, in both cases we have $p_1 = \cdots = p_4 = 0$.

Now, as $v(\boldsymbol{q}^4) = 2$, we have $y_5^4 + y_6^4 + y_7^4 = 2$, so at least one of the payoffs $y_5^4$, $y_6^4$ and $y_7^4$ is strictly positive. Assume without loss of generality that $y_5^4 = \delta > 0$. Then players 6, 7 and their partners in $\boldsymbol{q}^2$ and $\boldsymbol{q}^3$ (i.e., players $i'$, $i''$ such that $q_{i'}^2 = 1$, $q_{i''}^3 = 1$) can r-profitably deviate from $(CS', \boldsymbol{y})$ by forming a coalition structure $CS'' = (\boldsymbol{s}^1, \boldsymbol{s}^2, \boldsymbol{s}^3)$, where $\boldsymbol{s}^1$ is given by

$$s_{i'}^1 = 1, \quad s_6^1 = \frac{2}{3}, \quad s_\ell^1 = 0 \text{ for } \ell \neq i', 6,$$

$\boldsymbol{s}^2$ is given by

$$s_{i''}^2 = 1, \quad s_7^2 = \frac{2}{3}, \quad s_\ell^2 = 0 \text{ for } \ell \neq i'', 7,$$

and $\boldsymbol{s}^3 = (0, 0, 0, 0, 0, \frac{1}{3}, \frac{1}{3})$. We will now construct an imputation $\boldsymbol{z}$ for $CS''$ by setting $z_{i'}^1 = z_{i''}^2 = \frac{\delta}{4}$, $z_6^1 = z_7^2 = 100 - \frac{\delta}{4}$, $z_6^3 = y_6^4 + \frac{\delta}{2}$, $z_7^3 = y_7^4 + \frac{\delta}{2}$, and $z_i^j = 0$ for all $(i, j) \neq (i', 1), (6, 1), (i'', 2), (7, 2), (6, 3), (7, 3)$. It is not hard to see that $\boldsymbol{z} \in \mathcal{I}(CS'')$, and, moreover, the deviation $(CS'', \boldsymbol{z})$ is r-profitable for 6, 7, $i'$ and $i''$. Hence, $(CS', \boldsymbol{y})$ is not in the r-core of $G$. □

**Proposition 2.** *There exists an OCF-game $G$ such that* r-core$(G) \neq \emptyset$ *while* o-core$(G) = \emptyset$.

*Proof.* Consider an OCF-game $G = (N, v)$ with 3 agents $N = \{1, 2, 3\}$ whose weights are given by $\boldsymbol{w} = (8, 8, 8)$, and 2 task types $t^1$ and $t^2$. The first task needs 6 units of weight from each player, and has value 300, i.e. $v(r_1, r_2, r_3) = 300$ if $w_i r_i \geq 6$ for $i = 1, 2, 3$. The second task needs 4 units of weight in total from any of the players and has value 2.

Let $CS = (\boldsymbol{r}^1, \boldsymbol{r}^2)$, where $\boldsymbol{r}^1 = \left(\frac{7}{8}, \frac{7}{8}, \frac{6}{8}\right)$, $\boldsymbol{r}^2 = \left(\frac{1}{8}, \frac{1}{8}, \frac{2}{8}\right)$. Clearly, $v(\boldsymbol{r}^1) = 300$, $v(\boldsymbol{r}^2) = 2$. Consider also an imputation $\boldsymbol{x} \in \mathcal{I}(CS)$ given by $\boldsymbol{x}^1 = (100, 100, 100)$, $\boldsymbol{x}^2 = (0.5, 0.5, 1)$. It is not hard to see that $(CS, \boldsymbol{x}) \in$ r-core$(G)$. Indeed, as $CS$ maximizes the social welfare, there is no deviation that will be simultaneously profitable for all agents. Furthermore, if any agent withdraws his contribution from $\boldsymbol{r}^1$, he will lose the associated payoff of 100 and no deviation can compensate for this loss. Moreover, it is clear that withdrawing contribution from $\boldsymbol{r}^2$ cannot be profitable either, as there is no way to earn more than $2 = v(\boldsymbol{r}^2)$ with this amount of weight.

We will now show that $G$ has an empty o-core. Suppose for the sake of contradiction that there exists an outcome $(CS', \boldsymbol{y}) \in$ o-core$(G)$. It is not hard to show that any outcome in the o-core of $G$ maximizes the social welfare; the proof is similar to that of Theorem 4. Hence, we can assume that $CS' = (\boldsymbol{q}^1, \boldsymbol{q}^2)$, where $v(\boldsymbol{q}^1) = 300$, $v(\boldsymbol{q}^2) = 2$, and, moreover, $q_i^1 \geq \frac{6}{8}$ for $i = 1, 2, 3$. We have $y_1^2 + y_2^2 + y_3^2 = 2$, so we can assume without loss of generality that $y_1^2 = \delta > 0$. This means that players 2 and 3 can o-profitably deviate from $(CS', \boldsymbol{y})$ as follows: players 2 and 3 withdraw $q_2^1 w_2 - 6$ and $q_3^1 w_3 - 6$ units of weight from $\boldsymbol{q}^1$, respectively (as argued above, we have $q_2^1 w_2 \geq 6$, $q_3^1 w_3 \geq 6$), as well as their entire contribution to $\boldsymbol{q}^2$, and use these resources to complete $t^2$. If they divide the resulting payoff by allocating $y_2^2 + \frac{\delta}{2}$ to player 2 and $y_3^2 + \frac{\delta}{2}$ to player 3, this constitutes an o-profitable deviation for them. Thus, $(CS', \boldsymbol{y})$ is not in the o-core of $G$. $\qquad\square$

Thus, so far in this section we investigated the relationships among our notions of the overlapping core; it is also insightful to compare them to the non-overlapping and the fuzzy one. We now proceed to do so.

## 8.1 Comparison with the Non-Overlapping Core

Given an OCF-game $G = (N, v)$, we can define a non-overlapping game $G^{no} = (N, v^{no})$ by setting $v^{no}(C) = v(\boldsymbol{r}^C)$, where the partial coalition $\boldsymbol{r}^C$ is given by $r_i^C = 1$ if $i \in C$ and $r_i^C = 0$ otherwise for all $C \subseteq N$. Observe that for a threshold task game $G$ applying this transformation to its overlapping version $\check{G}$ gives us exactly its non-overlapping version $\hat{G}$. We can now compare the core of the game $G^{no}$ and the overlapping cores of the original game $G$. In particular, it is natural to ask whether the core of $G^{no}$ can be empty when the o-core of $G$ (and hence by Theorem 5 also the r-core and the c-core of $G$) is not, and vice versa, i.e., whether the c-core (the largest of the overlapping cores) of $G$ can be empty when the core of $G^{no}$ is not. Interestingly, it turns out that the answer to both of these questions is positive. We demonstrate this via examples based on threshold task games; as argued above, for any such game $G$ we have $\check{G}^{no} = \hat{G}$.

**Proposition 3.** *There exists a TTG $G$ with core$(\hat{G}) = \emptyset$, but o-core$(\check{G}) \neq \emptyset$.*

*Proof.* Consider a threshold task game $G = (N; \boldsymbol{w}; \boldsymbol{t})$, where $N = \{1, 2, 3\}$, $\boldsymbol{w} = (2, 2, 2)$, $\boldsymbol{t} = t^1 = (3, 1)$. In $\hat{G}$, any coalition structure $CS$ contains at most one coalition $C$ with $v(C) = 1$. Let $\boldsymbol{p} = (p_1, p_2, p_3)$ be an imputation for $CS$. As $v(CS) = 1$, there exists some $i \in N$ with $p_i > 0$. Then the coalition $C' = N \setminus \{i\}$ can successfully deviate from $(CS, \boldsymbol{p})$, as we have $w(C') = 4$, $p(C') = 1 - p_i < 1$. Hence, any outcome of $\hat{G}$ is not stable.

In $\check{G}$, the players can form two successful partial coalitions. Now, consider an outcome $(CS, \boldsymbol{x})$, where $CS = (\boldsymbol{r}^1, \boldsymbol{r}^2)$ with $\boldsymbol{r}^1 = (1, \frac{1}{2}, 0)$, $\boldsymbol{r}^2 = (0, \frac{1}{2}, 1)$, and $\boldsymbol{x}^1 = (\frac{2}{3}, \frac{1}{3}, 0)$, $x^2 = (0, \frac{1}{3}, \frac{2}{3})$. We claim that $(CS, \boldsymbol{x})$ is in the o-core of $\check{G}$. Indeed, suppose for the sake of contradiction that there is a group of players $J$ that has an o-profitable deviation from $(CS, \boldsymbol{x})$. We have $|J| \in \{1, 2, 3\}$. It is easy to see that $|J| \neq 1$: no player has enough weight to complete $t^1$ on his own. Also, $|J| \neq 2$: any pair of players earns $\frac{4}{3}$ in $(CS, \boldsymbol{x})$, and on their own they can make at most $1 < \frac{4}{3}$. Finally, $|J| \neq 3$, as $(CS, \boldsymbol{x})$ maximizes the social welfare. The contradiction completes the proof. $\qquad\square$

Intuitively, Proposition 3 holds because $\check{G}$ has more feasible outcomes than $\hat{G}$, and some of these additional outcomes turn out to be stable. On the flip side, $\check{G}$ allows for a wider range of deviations, so an outcome that is stable with respect to $\hat{G}$ may be unstable with respect to $\check{G}$. Our next proposition illustrates this.

**Proposition 4.** *There exists a TTG $G$ with* c-core$(\check{G}) = \emptyset$, *but* core$(\hat{G}) \neq \emptyset$.

*Proof.* Consider a threshold task game $G = (N; \boldsymbol{w}; \boldsymbol{t})$, where $N = \{1, 2, 3\}$, $\boldsymbol{w} = (9, 1, 1)$, $\boldsymbol{t} = (t^1, t^2)$ with $t^1 = (8, 100)$, $t^2 = (2, 1)$.

In $\hat{G}$, player 1 can work on task $t^1$, while players 2 and 3 can cooperate on task $t^2$, sharing the profits equally. Clearly, the resulting outcome is stable.

On the other hand, $\check{G}$ has no c-stable outcomes. Indeed, suppose that there is an outcome $(CS, \boldsymbol{x})$ in the c-core of $\check{G}$, and let $\boldsymbol{p}$ be the corresponding payoff vector. By Theorem 4, $CS$ consists of two partial coalitions: $\boldsymbol{r}^1$, which completes $t^1$, and $\boldsymbol{r}^2$, which completes $t^2$. Hence, $v(CS) = 101$. If $p_1 > 100$, then $p_2 + p_3 < 1$, and hence players 2 and 3 can deviate by forming a coalition $\boldsymbol{r} = (0, 1, 1)$ that can complete $t^2$ and has value 1. If $p_1 < 100$, player 1 can deviate by forming a coalition $\boldsymbol{r} = (1, 0, 0)$ that can complete $t^1$ and has value 100. Hence, we have $p_1 = 100$, $p_2 + p_3 = 1$, and therefore we can assume without loss of generality that $p_2 \leq \frac{1}{2}$. Now, players 1 and 2 can deviate by forming a coalition structure $CS' = (\frac{8}{9}, 0, 0), (\frac{1}{9}, 1, 0)$ and distributing the payoffs as $((100, 0, 0), (\frac{1}{3}, \frac{2}{3}, 0))$. We conclude that $(CS, \boldsymbol{x})$ is not c-stable, a contradiction. $\quad\square$

## 8.2 Comparison with Fuzzy Games

As mentioned earlier in this paper, Aubin (1981) introduces the notion of a *fuzzy game*, in which a player can participate in a coalition at various levels, and the value of a coalition $S$ depends on the participation levels of its members. Thus, at a first glance, the definition of a fuzzy game is identical to the definition of an OCF-game, as both are given by characteristic functions defined on $[0, 1]^n$. However, there are several crucial differences between fuzzy and OCF-games.

First, fuzzy games and OCF-games differ in their definition of an outcome. Indeed, while in OCF-games an outcome is an (overlapping) coalition structure together with a list of payoff vectors, in fuzzy games the only allowable outcome is the formation of the grand coalition. Furthermore, an outcome of an OCF-core needs to be stable against any deviation of a set $S$ to a (possibly overlapping) coalition structure. In the Aubin core, outcomes need only be stable against a deviation to a partial ("fuzzy") coalition, but not necessarily against deviations to a coalition structure. Indeed, the formation of coalition structures (overlapping or not) is not addressed in the fuzzy games literature.

One could try to represent games with overlapping coalition structures using the fuzzy games formalism. Indeed, given an OCF-game, we can construct a fuzzy game whose characteristic function simulates the behaviour of the characteristic function of the original OCF-game on coalition structures. Specifically, given any OCF-game $G = (N, v)$, we define a related fuzzy game $G' = (N, v')$ as follows. For any $\boldsymbol{r} \in [0, 1]^n$, we define

$$\mathcal{CS}_{\boldsymbol{r}} = \{(\boldsymbol{q}^1, \ldots, \boldsymbol{q}^k) \mid k \geq 1, q_i^j \geq 0 \text{ for } i = 1, \ldots, n, j = 1, \ldots, k, \sum_{j=1}^{k} q_i^j = r_i\},$$

and set $v'(\boldsymbol{r}) = \sup_{CS \in \mathcal{CS}_{\boldsymbol{r}}} v(CS)$. That is, for each partial coalition $\boldsymbol{r}$, $v'$ identifies the best coalition structure $CS$ that can be obtained by splitting $\boldsymbol{r}$ into subcoalitions, and returns its value $v(CS)$. The resulting fuzzy game $G'$ is very similar to the original OCF-game $G$. For example, for TTGs, this transformation would enable the members of the grand coalition to work on several tasks simultaneously. More generally, given a TTG $G$, any outcome of $(\check{G})'$ (i.e., a payoff vector for the grand coalition) corresponds to a social-welfare maximizing outcome $(CS, \boldsymbol{x})$ of $\check{G}$ and vice versa.

In fact, this relationship holds between any OCF-game $G$ and the corresponding fuzzy game $G'$ as long as the set $\{v(CS) \mid CS \in \mathcal{CS}_{(1,\dots,1)}\}$ is compact (and thus contains its least upper bound).

However, this approach fails to capture several delicate aspects of overlapping coalition formation. The main reason for this is that in the fuzzy game formulation, the actual set of tasks executed by a partial coalition is implicit in the definition of the characteristic function. Indeed, an outcome of the fuzzy game is simply a payoff vector, and while we are guaranteed that there is a set of tasks that provides the corresponding total payoff, this set of tasks cannot be "read off" the description of the outcome. This leads to a number of difficulties.

First, the fuzzy games formalism would not allow us to reason about partial coalition structures with suboptimal social welfare. While by Theorem 4 such coalition structures are unlikely to be the final outcomes of a game, a dynamic coalition formation protocol may produce such partial coalition structures as intermediate steps. Thus, using the language of fuzzy coalitions impairs our ability to study the processes that lead to the formation of partial coalition structures. As such processes are of great interest from the practical perspective, this is an important disadvantage of the fuzzy model.

Further, under the OCF representation, there is a one-to-one correspondence between partial coalitions and tasks. This makes the OCF approach intuitively appealing, and suggests that it provides the right level of granularity for reasoning about partial coalition formation. Indeed, consider our problem from a computational perspective in the context of TTGs. While under the OCF representation finding a socially optimal coalition structure can be difficult (see Appendix A), computing the value of a given partial coalition $\boldsymbol{r}$ is straightforward: we simply pick the most valuable task that can be completed using the resources possessed by $\boldsymbol{r}$. In contrast, in the fuzzy game framework, the two issues are intertwined, so even computing a partial coalition's worth is a hard problem.

Even more importantly, the definition of the fuzzy core given by Aubin (1981) is not appropriate for many natural scenarios, and, in particular, TTGs. Specifically, the fuzzy core of a fuzzy game $G = (N, v)$ is defined as the set of all outcomes $(N, \boldsymbol{p})$ such that $p(N) = v(1, \dots, 1)$ and for any partial coalition $\boldsymbol{r}$ it holds that $\sum_{i=1}^{n} p_i r_i \geq v(\boldsymbol{r})$. Essentially, this means that when a group of players deviates from the grand coalition via a partial coalition $\boldsymbol{r}$, each deviating player $i$ receives both her payoff from $\boldsymbol{r}$, and her original payoff from the grand coalition, scaled down by a factor of $(1 - r_i)$. Thus, the fuzzy core is even more "optimistic" from the deviators' perspective than the o-core. Indeed, the deviators do not worry about what the grand coalition will be able to do once they leave. They simply assume that if they withdraw, say, 40% of their resources, they will get 60% of what they used to get. However, in many games—and, in particular, TTGs—if some players abandon the grand coalition, the latter may not have sufficient resources to complete any task. Clearly, in this case the deviators could not possibly get any payoff from what remains of the grand coalition. Thus, the fuzzy core may be empty, even if in practice the game is stable. The example in the proof of Proposition 5 illustrates this.

**Proposition 5.** *There exists a TTG $G$ such that* o-core$(\check{G}) \neq \emptyset$, *but the fuzzy core of the corresponding fuzzy game $(\check{G})'$ is empty.*

*Proof.* Consider a TTG $G$ given by $N = \{1, 2\}$, $\boldsymbol{w} = (10, 10)$, and $\boldsymbol{t} = ((20, 20), (7, 9))$, and the induced OCF-game $\check{G}$. The corresponding fuzzy game $(\check{G})' = (N, v')$ is given by

$$v'(\boldsymbol{r}) = \begin{cases} 20 & \text{if } r_1 + r_2 = 2 \\ 18 & \text{if } 1.4 \leq r_1 + r_2 < 2 \\ 9 & \text{if } 0.7 \leq r_1 + r_2 < 1.4 \\ 0 & \text{if } r_1 + r_2 < 0.7 \end{cases}$$

It is not hard to see that the outcome $(CS, \boldsymbol{x})$ of $\check{G}$, where $CS = \boldsymbol{r} = (1, 1)$ and $\boldsymbol{x} = (10, 10)$ is o-stable. Moreover, intuitively, it is clear that no rational agent or a coalition of agents would want to deviate from this outcome. On the other hand, under the definition of the fuzzy core the outcome $(10, 10)$ of $(\hat{G})'$ is not stable: indeed, for $\boldsymbol{q} = (.7, .7)$ we have $p_1 q_1 + p_2 q_2 = 14 < 18 = v'(\boldsymbol{q})$.

We will now prove that *no* outcome of $(\check{G})'$ is in the fuzzy core. Observe that since $v'(1, 1) = 20$, any outcome of $(\hat{G})'$ is of the form $(z_1, z_2)$, where $z_1 + z_2 = 20$. Clearly, any outcome with $z_1 < 9$ or $z_2 < 9$ is unstable, as the partial coalition $(1, 0)$ (respectively, $(0, 1)$) can profitably deviate from it. Thus we can assume that $z_1 \geq 9$, $z_2 \geq 9$, or, equivalently, $z_2 \leq 11$, $z_1 \leq 11$. Thus, for the partial coalition $\boldsymbol{q}$ considered above, we have $z_1 q_1 + z_2 q_2 \leq 11 \times 1.4 = 15.4 < 18 = v(\boldsymbol{q})$, which means that $(z_1, z_2)$ is not in the fuzzy core. $\qquad \square$

**Remark 4.** *To remedy some of the difficulties illustrated above, we can devise a notion of stability that is defined within the framework of fuzzy games, yet is essentially equivalent to the c-core. Let us say that an outcome $\boldsymbol{p}$ of $G'$ is* f-stable *if for any $\boldsymbol{r} \in [0, 1]^n$ we have $v'(\boldsymbol{r}) \leq \sum_{i \in \text{supp}(\boldsymbol{r})} p_i$, and define the* f-core *of $G'$ to be the set of all f-stable outcomes of $G'$. Note that this definition is different from the standard definition of the fuzzy core. For TTGs, one can show that an outcome $\boldsymbol{p}$ of $G'$ is in the* f-core *of $G'$ if and only if the corresponding outcome $(CS, \boldsymbol{x})$ of $\check{G}$ is in the c-core of $\check{G}$. The proof makes use of the fact that in TTGs one can distribute the profit $v'(\boldsymbol{r})$ of a deviating partial coalition $\boldsymbol{r}$ among the members of $\text{supp}(\boldsymbol{r})$ arbitrarily. (In more detail, one can construct a partial coalition structure $CS$ involving agents in $\text{supp}(\boldsymbol{r})$ that performs tasks of total value $v'(\boldsymbol{r})$ so that each agent in $\text{supp}(\boldsymbol{r})$ participates in each partial coalition in $CS$.) Moreover, this equivalence is true for general OCF games whose characteristic functions satisfy some natural regularity conditions; the proof is similar to the proof of Theorem 1. Unfortunately, while the f-core provides an analogue of the c-core in the fuzzy game setup, it is not clear how to devise an analogue of the r-core or the o-core for this setting. Indeed, to define these concepts, we would have to reason about partial coalitions that are hurt by a deviation. However, the description of an outcome of a fuzzy game does not indicate which partial coalitions a given player belongs to, so we cannot determine which tasks will be affected by a deviation.*

We conclude that there are natural settings where OCF-games provide a more realistic and nuanced model than fuzzy games; threshold task games appear to be one such example.

## 9. Computational Aspects of Stability in Threshold Task Games

In this section, we investigate the computational complexity of core-related questions in TTGs. Our goal here is twofold. First, TTGs provide a natural model of agent collaboration, and therefore it is important to understand how to allocate resources in such games in a stable manner. Second, our analysis highlights important differences between the three definitions of the core for games with overlapping coalitions. In particular, the results presented in this section provide a complexity-theoretic separation between the c-core, on one hand, and the r-core and the o-core, on the other hand. We believe that results of this type are useful for building a better understanding of stability in the context of general OCF games.

Unless explicitly stated otherwise, we make the usual assumption that all parameters of the game—i.e., all weights, thresholds and task utilities—are integers given in binary. This assumption can be made without loss of generality, and is necessary for a formal complexity-theoretic analysis.

### 9.1 Games with Non-Overlapping Coalitions

We start by analyzing the complexity of TTGs in the non-overlapping setting. As mentioned in Section 5.1, such games can be seen as a generalization of weighted voting games with coalition structures. Elkind, Chalkiadakis & Jennings (2008) show that several stability-related questions in such games are computationally hard when weights are integers given in binary. Hence, we can formulate the following proposition, whose proof follows immediately from those results.

**Proposition 6.** *Given a TTG $G = (N; \boldsymbol{w}; \boldsymbol{t})$, it is* coNP-*hard to decide whether the corresponding game $\hat{G}$ has an empty core. Also, given an outcome $(CS, \boldsymbol{p})$ of $\hat{G}$, it is* coNP-*complete to decide whether $(CS, \boldsymbol{p})$ is in the core of $\hat{G}$. These results hold even if there is only one task type, and the utility of this task is $1$.*

On the other hand, Elkind et al. (2008) provide a polynomial-time algorithm for checking if an outcome of a weighted voting game is in the core if weights are given in unary. That algorithm is based on dynamic programming: given a weighted voting game $G$ described by a set of players $N$, a list of weights $\boldsymbol{w}$ and a threshold $T$, for each weight $1, \dots, w(N)$ it identifies the minimum payoff $P_w$ to a coalition that has weight $w$, and then checks if $P_w < 1$ for some $w \geq T$.

It is not hard to see that a similar approach works for threshold task games as well. The only complication is that for each weight $w$, in addition to computing the minimum payoff for a coalition of this weight under the given imputation, we have to compute the maximum utility available to a coalition of this weight, i.e., $\max\{u^j \mid w \geq T^j\}$, and compare the two quantities. However, these additional steps are very easy (in particular, they can be performed efficiently even if task utilities are large). This gives us the following result.

**Proposition 7.** *There exists an algorithm that, given a TTG $G = (N; \boldsymbol{w}; \boldsymbol{t})$ and an outcome $(CS, \boldsymbol{p})$ of $\hat{G}$, checks whether $(CS, \boldsymbol{p})$ is in the core of $\hat{G}$ and runs in time $\mathrm{poly}(w(N), |\boldsymbol{p}|)$, where $|\boldsymbol{p}|$ is the number of bits in the binary representation of $\boldsymbol{p}$.*

For weighted voting games with unary weights, Elkind et al. (2008) also show that, by constructing a linear program that uses the algorithm of Proposition 7 as an oracle, we can check in polynomial time whether a given coalition structure $CS$ can be stabilized, i.e., whether there exists a payoff vector $\boldsymbol{p} \in \mathcal{I}(CS)$ such that $(CS, \boldsymbol{p})$ is in the core. This algorithm can be easily adapted to work for TTGs with unary weights. Hence, the question of whether a given coalition structure can be stabilized is poly-time solvable for these games, too.

### 9.2 Games with Overlapping Coalitions

We will now show that, similarly to the non-overlapping case, if all weights, thresholds and utilities in a TTG are integers given in binary, then it is computationally hard to check if a given outcome of the corresponding OCF game is stable. Moreover, this hardness result holds for all three definitions of stabilty, i.e., the c-core, the r-core, and the o-core. While these results are perhaps not surprising given the similar result for the non-overlapping setting (i.e., Proposition 6 above), the reason behind the computational hardness is quite different. Indeed, the reduction used in the proof of Proposition 6 is based on PARTITION, a classic NP-hard problem which asks whether, given a set of weights, we can split it into two sets of the same weight. Essentially, the proof proceeds by constructing an outcome that is stable if and only if a certain subset of agents cannot be split into two groups that have the same weight. This proof technique is unlikely to work in the overlapping scenario, as one can always form two partial coalitions of the same weight by allowing all agents to split their weight equally between two coalitions. Hence, the proof of the following theorem uses a somewhat different approach.

**Theorem 6.** *Given a TTG $G = (N; \boldsymbol{w}; \boldsymbol{t})$ and an outcome $(CS, \boldsymbol{x})$ of the corresponding OCF game $\check{G}$, it is* coNP-*complete to decide whether $(CS, \boldsymbol{x})$ is in the c-core of $\check{G}$.*

*Proof.* Our reduction is based on UNBOUNDED KNAPSACK, a well-known NP-hard problem. An instance of UNBOUNDED KNAPSACK (Martello & Toth, 1990) is given by a set of $\ell$ items, where each item $i$ has a size $s_i$ and a value $z_i$, the knapsack size $B$ and the target value $Z$. It is a "yes"-instance if we can fill the knapsack using an unlimited number of copies of each item so that the total size of the resulting set of items is at most $B$, while their total value is at least $Z$, i.e., if there is a vector of non-negative integers $(\alpha_1, \ldots, \alpha_\ell)$ such that $\sum_{i=1}^{\ell} \alpha_i s_i \leq B$ and $\sum_{i=1}^{\ell} \alpha_i z_i \geq Z$. Otherwise, it is a "no"-instance.

Consider an instance $I = ((s_1, \ldots, s_\ell); (z_1, \ldots, z_\ell); B; Z)$ of UNBOUNDED KNAPSACK. We can assume without loss of generality that $s_j < B, z_j < Z$ for all $j = 1, \ldots, \ell$. Moreover, we can assume that $I$ is monotone, i.e., $s_i \leq s_j$ implies $z_i \leq z_j$. Indeed, if we have a pair of items such that $s_i \leq s_j$, but $z_i > z_j$, we can simply delete the $j$th item, as it is not used by any optimal solution.

We will now construct an instance of our problem as follows. Set $N = \{1\}$ and let $w_1 = B$. Set $\boldsymbol{t} = (t^1, t^2, \ldots, t^{\ell+1})$, where $T^j = s_j$, $u^j = z_j$ for $j = 1, \ldots, \ell$ and $T^{\ell+1} = B$, $u^{\ell+1} = Z - 1$. Due to our restrictions on $I$, the game $G = (N; \boldsymbol{w}; \boldsymbol{t})$ is a threshold task game.

Consider an outcome $(CS, \boldsymbol{p})$ where $CS$ consists of a single partial coalition $\boldsymbol{r}$ with $r_1 = 1$ and $\boldsymbol{p} \in \mathcal{I}(CS)$. As $B > s_j$ for all $j = 1, \ldots, \ell$, this coalition executes the task $t^{\ell+1}$ and receives utility of $Z - 1$. Hence, player 1 can c-profitably deviate from $(CS, \boldsymbol{p})$ if and only if he can find a collection of tasks whose total resource requirement is at most his weight $B$ and whose total utility is at least $Z$, i.e., if and only if we started with a "yes"-instance of UNBOUNDED KNAPSACK. □

In the proof of Theorem 6 the outcome $(CS, \boldsymbol{x})$ consists of a single partial coalition. Thus, any r-profitable deviation from $(CS, \boldsymbol{x})$ is c-profitable. This implies the following corollary.

**Corollary 2.** *Given a TTG $G$ and an outcome $(CS, \boldsymbol{x})$ of the corresponding OCF game $\check{G}$, it is* coNP-*complete to decide if $(CS, \boldsymbol{x})$ is in the r-core of $\check{G}$.*

For the o-core, the situation is somewhat more complicated. However, a more careful examination of the proof of Theorem 6 allows us to obtain the following corollary.

**Corollary 3.** *Given a TTG $G = (N; \boldsymbol{w}; \boldsymbol{t})$ and an outcome $(CS, \boldsymbol{x})$ of the corresponding OCF game $\check{G}$, it is* coNP-*complete to decide if $(CS, \boldsymbol{x})$ is in the o-core of $\check{G}$.*

*Proof.* In the proof of Theorem 6, we construct an OCF game with 1 player and an outcome $(\boldsymbol{r}, \boldsymbol{x})$. Consider any o-profitable deviation $(CS, \boldsymbol{y})$ from $(\boldsymbol{r}, \boldsymbol{x})$. This deviation itself is not necessarily a c-profitable deviation from $(\boldsymbol{r}, \boldsymbol{x})$: under $(CS, \boldsymbol{y})$, agent 1 may withdraw some, but not all of his resources from $(\boldsymbol{r}, \boldsymbol{x})$ and therefore continue to derive some benefit from it. However, for a single agent, allocating some of the resources to the original partial coalition $\boldsymbol{r}$ is equivalent to forming a new partial coalition using that amount of resources, i.e., given $(CS, \boldsymbol{y})$, one can construct a deviation from $(\boldsymbol{r}, \boldsymbol{x})$ that will be c-profitable for agent 1. On the other hand, any c-profitable deviation from $(\boldsymbol{r}, \boldsymbol{x})$ is also o-profitable. Hence, $(\boldsymbol{r}, \boldsymbol{x})$ is o-stable if and only if it is c-stable, i.e., if and only if we started with a "no"-instance of UNBOUNDED KNAPSACK. $\square$

In the rest of the section, we will focus on the case where all parameters of the game (i.e., all players' weights, all thresholds and all task utilities) are integers that are given in unary, or, equivalently, are at most polynomial in the number of players. Given a game $G = (N; \boldsymbol{w}; \boldsymbol{t})$, where $t^j = (T^j, u^j)$ for $j = 1, \ldots, m$, let $|G| = w(N) + \sum_{j=1}^{m}(T^j + u^j)$.

It turns out that in this setting checking whether an outcome is in the c-core becomes easy. Intuitively, the reason for this is that once a group of players decides to deviate, the agents in this group can easily decide how to proceed: they need to pool their weights and find the most profitable set of tasks that can be completed using this amount of resources.

**Theorem 7.** *There exists an algorithm that, given a TTG $G = (N; \boldsymbol{w}; \boldsymbol{t})$ and an outcome $(CS, \boldsymbol{x})$ of the corresponding OCF game $\breve{G}$, checks whether $(CS, \boldsymbol{x})$ is in the c-core of $\breve{G}$ and runs in time $\mathrm{poly}(|G|, |\boldsymbol{x}|)$, where $|\boldsymbol{x}|$ is the size of the binary representation of the imputation $\boldsymbol{x}$.*

*Proof.* Our algorithm is based on dynamic programming. First, for any $w = 1, \ldots, w(N)$, let $U_w$ be the maximum profit that a coalition of weight $w$ can make, i.e.,

$$ U_w = \max \left\{ \sum_{j=1}^{m} \alpha^j u^j \mid \sum_{j=1}^{m} \alpha^j T^j \leq w, (\alpha^1, \ldots, \alpha^m) \in \mathbb{N}^m \right\}. $$

For each $w = 1, \ldots, w(N)$, the quantity $U_w$ can be computed using the dynamic programming algorithm for UNBOUNDED KNAPSACK. The running time of this procedure is polynomial in $|G|$.

Now, let $\boldsymbol{p}$ be the payoff vector that corresponds to the imputation $\boldsymbol{x}$. For all $i = 1, \ldots, n$ and all $w = 1, \ldots, w(N)$, set $P_{i,w} = \min\{p(S) \mid S \subseteq \{1, \ldots, i\}, w(S) = w\}$. The quantities $P_{i,w}$ can be easily computed using dynamic programming. Indeed, we have $P_{1,w} = p_1$ if $w = w_1$ and $P_{1,w} = +\infty$ otherwise (we use the convention that $\min \emptyset = +\infty$). Furthermore, we can compute $P_{i+1,w}$ given the values $(P_{i,w'})_{w'=1,\ldots,w}$ by setting $P_{i+1,w} = \min\{P_{i,w}, p_i + P_{i,w-w_i}\}$. The running time of this procedure is $\mathrm{poly}(|G|, |\boldsymbol{p}|)$.

Suppose that we have computed $P_{n,w}$ for $w = 1, \ldots, w(N)$. Observe that the value $P_{n,w}$ is the least amount received by a coalition of weight $w$ under $\boldsymbol{p}$. Now, for each $w = 1, \ldots, w(N)$, we can compare the quantities $P_{n,w}$ and $U_w$. If there is a value of $w$ for which the latter exceeds the former, there is a coalition in $N$ that could increase its collective earnings by deviating from $(CS, \boldsymbol{x})$, i.e., $(CS, \boldsymbol{x})$ is not in the c-core of $\breve{G}$. It is not hard to see that the converse is also true: if $P_{n,w} \geq U_w$ for all $w = 1, \ldots, w(N)$, then no coalition has a c-profitable deviation from $(CS, \boldsymbol{x})$, and hence $(CS, \boldsymbol{x})$ is in the c-core of $\breve{G}$.

Clearly, this algorithm runs in time $\mathrm{poly}(|G|, |\boldsymbol{x}|)$. $\square$

In contrast, the corresponding problems for the r-core and the o-core are computationally hard. Intuitively, the reason for this is that the decisions the players make are no longer binary: instead of simply deciding whether or not to deviate, they have to decide which of their coalitions with non-deviators to abandon. In the case of the o-core, there is also the possibility of reducing one's contribution to a partial coalition rather than abandoning it altogether.

**Theorem 8.** *Given a TTG $G = (N; \boldsymbol{w}; \boldsymbol{t})$ and an outcome $(CS, \boldsymbol{x})$ of the corresponding OCF game $\check{G}$, it is strongly* coNP-*complete to decide whether $(CS, \boldsymbol{x})$ is in the r-core of $\check{G}$.*

*Proof.* It is not hard to see that this problem is in coNP: to show that an outcome $(CS, \boldsymbol{x})$ is not in the r-core of $\check{G}$, we can guess a set of deviators $J$ and a deviation $(CS', \boldsymbol{y})$, and check that $(CS', \boldsymbol{y})$ is r-profitable for $J$ by computing the payoffs of all players in $J$ under $\boldsymbol{x}$ and $\boldsymbol{y}$.

To show coNP-hardness, we reduce from MAXIMUM EDGE BICLIQUE (Peeters, 2003). An instance on MAXIMUM EDGE BICLIQUE is given by a bipartite graph $B = (L, R, E)$ with a set of vertices $L \cup R$ and a set of edges $E \subseteq L \times R$, and a parameter $K$. It is a "yes"-instance if $B$ contains a biclique of size at least $K$, i.e., if there are sets $L' \subseteq L$, $R' \subseteq R$ such that $|L'| * |R'| \geq K$, and for all $\lambda \in L$ and all $\rho \in R$ we have $(\lambda, \rho) \in E$. Otherwise, it is a "no"-instance.

Suppose that we are given an instance $(B, K)$ of MAXIMUM EDGE BICLIQUE with $B = (L, R, E)$, $L = \{\lambda_1, \ldots, \lambda_{|L|}\}$, $R = \{\rho_1, \ldots, \rho_{|R|}\}$. Then we create an instance of our problem as follows. Assume without loss of generality that $|L| \leq |R|$. We set $n = |R| + 1$, $k = |L|$, $M = k^2 n^2$, $V = k^2 n M$, and create $n$ players with weights $w_1 = \cdots = w_{n-1} = k$, $w_n = k(kn - n + 1)$ and 2 task types $t^1 = (kn; V)$ and $t^2 = (K; (n-1)k+1)$. Also, we create a coalition structure $CS = (\boldsymbol{r}^1, \ldots, \boldsymbol{r}^k)$ given by $r_i^j = 1/k$ for all $i = 1, \ldots, n$ and all $j = 1, \ldots, k$. Observe that the total weight of each $r^j \in CS$ is $kn$, so each such partial coalition performs $t^1$. Finally, to construct the imputation $\boldsymbol{x} = (\boldsymbol{x}^1, \ldots, \boldsymbol{x}^k)$, for all $j = 1, \ldots, k$ and all $i = 1, \ldots, n-1$, we set $x_i^j = 1$ if $(i, j) \in E$ and $x_i^j = M$ otherwise. Also, we set $x_n^j = V - \sum_{i=1}^{n-1} x_i^j$ for all $j = 1, \ldots, k$.

Suppose we started with a "yes"-instance of MAXIMUM EDGE BICLIQUE, and let $(L', R')$ be the corresponding subgraph of $B$. Then the subset of players $J = \{i \mid \rho_i \in R'\}$ can r-profitably deviate from $(CS, \boldsymbol{x})$ by abandoning the partial coalitions in the set $S = \{\boldsymbol{r}^j \mid \lambda_j \in L'\}$, and using the freed-up resources to embark on $t^2$. Indeed, under $\boldsymbol{x}$ the players in $J$ collectively earn at most $(n-1)k$ from partial coalitions in $S$, and devote at least $K$ units of weight to these coalitions.

Conversely, consider any r-profitable deviation $(CS', \boldsymbol{y})$, and let $J$ be the corresponding set of deviators. Suppose that $k_1$ coalitions in $CS'$ work on $t^1$, and $k_2$ coalitions work on $t^2$. First, suppose $n \in J$. Observe that $(CS', \boldsymbol{y})$ is profitable for player $n$ if and only if $k_1 = k$, $k_2 = 0$: indeed, under $(CS, \boldsymbol{x})$ player $n$ earns at least $k(V - (n-1)M)$, whereas under any outcome that completes less that $k$ copies of $t^1$ he earns at most $(k-1)V + \frac{k^2 n}{K}((n-1)k+1) < k(V - (n-1)M)$. However, any deviation that results in executing $k$ copies of $t^1$ must involve all resources of all players, i.e., $J = \{1, \ldots, n\}$, and any such deviation cannot be simultaneously profitable for all members of the deviating set. Hence, we have $n \notin J$, and therefore $w(J) \leq k(n-1)$. Consequently, $k_1 = 0$ and the deviators' total profit is at most $\frac{w(J)}{K}((n-1)k+1) < M$. This means that $(CS', \boldsymbol{y})$ is an r-profitable deviation only if no player $i \in J$ abandons a coalition $\boldsymbol{r}^j \in CS$ such that $x_i^j = M$. On the other hand, to successfully execute even one copy of $t^2$, the members of $J$ must collectively withdraw at least $K$ units of weight. Let $R' = \{\rho_i \mid i \in J\}$, and let $L'$ correspond to the set of partial coalitions in $CS$ affected by the deviation; then $(L', R')$ is a biclique of size at least $K$. $\square$

It is not hard to check that in the proof of Theorem 8 no player can withdraw part of his resources from a partial coalition in $CS$ and still claim any profit from that coalition. This implies that checking whether a given outcome is in the o-core is computationally hard, too.

**Corollary 4.** *Given a TTG $G$ and an outcome $(CS, \boldsymbol{x})$ of the corresponding OCF game $\check{G}$, it is strongly* coNP-*complete to decide whether $(CS, \boldsymbol{x})$ is in the o-core of $\check{G}$.*

On the other hand, combining the techniques of Theorem 7 and Theorem 4 leads to a pseudopolynomial algorithm for checking whether the c-core of a TTG is non-empty.

**Theorem 9.** *Given a TTG $G = (N; \boldsymbol{w}; \boldsymbol{t})$, one can check in time $\mathrm{poly}(|G|)$ whether the corresponding OCF game $\check{G}$ has a non-empty c-core.*

*Proof.* We will show that if the c-core of a game $\check{G}$ is non-empty, then for any social welfare-maximizing set of tasks we can construct a coalition structure $CS$ that executes this set of tasks and an imputation $\boldsymbol{x} \in \mathcal{I}(CS)$ such that $(CS, \boldsymbol{x})$ is in the c-core of $\check{G}$; moreover, in $CS$ each agent contributes to each coalition. Hence, our algorithm first selects a social welfare-maximizing set of tasks, then constructs a coalition structure that can perform this set of tasks, and finally solves a linear program to check if this coalition structure can be stabilized. The details follow.

Assume for simplicity that $\boldsymbol{t}$ contains a task type $t$ with $T = 1$; if this is not the case we can add a task type $t^0 = (1, 0)$ to $\boldsymbol{t}$. This allows us to assume that in any coalition structure all agents' resources are committed to some tasks. Fix a social welfare-maximizing multi-set of tasks $\{\alpha_1 t^1, \ldots, \alpha_m t^m\}$. Suppose c-core$(\check{G}) \neq \emptyset$, and let $(CS', \boldsymbol{y})$ be an outcome in the c-core of $\check{G}$. By Theorem 4, we have $\sum_{j=1}^m \alpha_j u^j = v(CS')$. Consider a coalition structure $CS$ that contains $\alpha_1 + \cdots + \alpha_m$ coalitions: the first $\alpha_1$ coalitions have weight $T^1$ each, the next $\alpha_2$ coalitions have weight $T^2$ each, etc., and each agent $i$ distributes his resources evenly between all coalitions, i.e., he contributes $w_i \frac{T^1}{w(N)}$ units of weight to each of the first $\alpha_1$ coalitions, etc. As in $CS$ all agents contribute to all partial coalitions, and $v(CS) = v(CS')$, we have $\boldsymbol{y} \in \mathcal{I}(CS)$. Moreover, it is clear that the outcome $(CS, \boldsymbol{y})$ is in c-core$(\check{G})$: any c-profitable deviation from $(CS, \boldsymbol{y})$ is also a c-profitable deviation from $(CS', \boldsymbol{y})$.

By Proposition 9 when all weights are given in unary, we can find a social welfare-maximizing coalition structure $CS = (\boldsymbol{r}^1, \ldots, \boldsymbol{r}^k)$ in polynomial time. Consider the following linear program:

$$\begin{aligned} p_i &\geq 0 \text{ for } i = 1, \ldots, n \\ \sum_{i \in N} p_i &= v(CS) \\ \sum_{i \in J} p_i &\geq U_{w(J)} \text{ for all } J \subseteq N, \end{aligned}$$

where $U_w$ is defined as in the proof of Theorem 7. While this linear program has exponentially many constraints, it can be solved in linear time by the ellipsoid method (Schrijver, 1986), since it has a polynomial-time separation oracle. Indeed, we can decide whether a given candidate solution is feasible using the algorithm described in the proof of Theorem 7.

Clearly, if this linear program has a feasible solution $\boldsymbol{p}$, then the imputation $\boldsymbol{x}$ given by $x_i^j = p_i \frac{v(\boldsymbol{r}^j)}{v(CS)}$ for all $i \in N$ and all $j = 1, \ldots, |CS|$ satisfies $\boldsymbol{x} \in \mathcal{I}(CS)$, and, moreover, $(CS, \boldsymbol{x}) \in$ c-core$(\check{G})$. Conversely, if it does not have a feasible solution, then $CS$ cannot be stabilized, and hence by the argument above the c-core of $\check{G}$ is empty. □

## 10. Conclusions, Extensions, and Future Work

In this paper we introduced a model of cooperative games that allows for overlapping coalitions and takes into account the need for resource allocation. In doing so, we generalized the usual models where either the grand coalition is the only desirable outcome or the outcomes are required to be partitions of the set of agents. Given our model, we defined and studied in depth a notion of the core (the *c-core*) which is a generalization of the core in the traditional models of cooperative game theory. Under some quite general conditions, we provided a characterization for an outcome—that is, a *(coalition structure, imputation)* pair—to belong to the core. We also showed that any outcome in the core maximizes the social welfare. Further, we introduced a notion of balancedness for OCF-games, and showed that a coalition structure $CS$ admits an imputation $\boldsymbol{x}$ so that $(CS, \boldsymbol{x})$ is in the core if and only if the game is balanced. Moreover, we extended the notion of convexity to our setting and showed that convex games have a non-empty core.

In addition, we considered two other notions of core-stability in OCF-games, which differ from each other (as well as from the first one) in what the deviators expect to obtain from their collaboration with non-deviators. Together, our three notions of the core span a wide range of beliefs that the deviators may hold regarding payoffs from coalitions with non-deviators, and can be substantially different from each other with respect to the sets of outcomes that they characterize, and with respect to their computational complexity. We also compared the OCF-games with their non-overlapping analogues, and showed that from the social welfare maximization perspective, OCF-games may provide higher total utility, and are easier to work with than their classic counterparts. We have also argued that OCF-games provide a more appropriate modelling framework than fuzzy games for many scenarios; in particular, this is certainly the case for threshold task games. To summarize, our paper is one of the very first attempts to provide a theoretical treatment of overlapping coalition formation, and to study stability in this setting in a thorough manner.

### 10.1 Extensions

In many environments, when a coalition is formed, it may have a choice of actions to execute. While in a deterministic setting such as the one considered in this paper, the coalition will simply choose the action that results in the highest possible payoff, in a probabilistic environment this choice is more difficult: a coalition may want to strike a balance between the expected payoff and the variance. To address this issue, we can incorporate *coalitional actions* in our model as follows.

A coalition is allowed to select an action from a (usually finite) action space $\mathcal{A}$. Without loss of generality, we assume that each coalition can undertake any action in $\mathcal{A}$.[3] The value of a coalition is then determined by the resource contribution levels of its members *and* the action selected. Therefore, the characteristic function in our setting is then defined on $(\boldsymbol{r}, a)$ pairs, where $\boldsymbol{r} = (r_1, \ldots, r_n)$ is a vector of resources, and $a \in \mathcal{A}$ is an action. All of our definitions and results generalize readily to the situation where each coalition has a choice of actions (simply put, our presentation so far corresponds to a situation where each coalition had exactly one action available to it).

Another extension we have examined has to do with modelling the available resources. For ease of presentation it was assumed throughout the paper that there exists only one type of (continuous) resource. Nevertheless, all of our results still hold if we assume multiple types of resources (e.g., agents have to distribute both time *and* money among their coalitions). Moreover, we have also studied a "discrete" OCF setting, with agent contribution levels taking values in a finite set (i.e., an agent may be able to contribute 20%, but not 21% of his resources to a coalition). Such a setting is obviously of interest in many applications involving countable resources (as the discretization of effectively any kind of resources is common in practice). With discrete resources, the number of possible coalition structures is now finite (as a coalition in our setting is a collection of resources—see Section 4). All of our definitions and theorems carry through in this setting with minor differences in the arguments used in the proofs.

---

3. The situation where this is not the case can be modeled by setting the value of the respective *(coalition, action)* pair to 0.

## 10.2 Future Work

There exist many exciting open questions for future work. First of all, an important research direction is to develop a better understanding of scenarios where overlapping coalitions can naturally arise, and to identify the appropriate stability concepts for these scenarios. We believe that techniques developed in this paper will prove useful for this purpose. Moreover, one of our first priorities is to investigate further the alternative notions of stability (i.e., the o-core and the r-core) proposed above, and obtain relevant characterization results, as we did with the c-core. Extending other solution concepts for coalitional games—such as, e.g., the Shapley value—to OCF settings is an important research direction as well.

We also plan to study further the computational complexity of core-related questions in this setting. First, while we have initiated the study of complexity-theoretic aspects of stability in OCF games, in this paper we have focused on the complexity of checking whether a given outcome is stable. Another natural problem in this domain is studying the complexity of checking whether a game has a stable solution—i.e., whether its c-core (r-core, o-core) is non-empty. Theorem 9 makes the first steps in this direction, suggesting that this problem may be easier in the overlapping setting than in the classic setting: indeed, Elkind et al. (2008) conjecture that for WVGs with coalition structures checking the non-emptiness of the core is hard for unary weights.

Now, the hardness results for computing an allocation in the core or checking if the core is non-empty in the traditional setting—as those in the work of Chvatal (1978), Tamir (1991), Deng and Papadimitriou (1994), Sandholm et al. (1999), Conitzer and Sandholm (2006)—and our hardness results in this paper suggest that one can only hope to identify special classes of games where we can have efficient algorithms for computing core allocations. As noted earlier, an element of the core in convex games can be computed in the traditional setting simply by taking the vector of the marginal contributions of the agents for an arbitrary permutation of the set of agents. In our setting, even though our proof yields a procedure for constructing an element of the c-core, it requires solving a series of optimization questions, which for arbitrary convex games are NP-hard. Naturally, it would be desirable to find classes of convex games where our proof yields a polynomial time algorithm.

We are also interested in finding processes that lead to the core in not necessarily convex games; though *randomized algorithms* such as the ones of Dieckmann and Schwalbe (1998) and Chalkiadakis and Boutilier (2004) trivially extend to the overlapping setting, they would be of little practical value here due to the huge space of potential overlapping configurations. Therefore, we are interested in finding ways to exploit known game structure to prune the search space for potential stable configurations. Another subject of future research is extending our model to allow for infinite coalition structures. Furthermore, it would be interesting to establish links between outcomes in the core and outcomes of bargaining equilibria in overlapping coalitional bargaining games.

Finally, the incorporation of actions in our model allows for the investigation of action stochasticity and, more generally, uncertainty in an OCF setting. For instance, a coalitional action can be associated with a distribution over possible payoff outcomes resulting from its execution. This poses challenges to study such models from both a theoretical and a practical standpoint, since the introduction of uncertainty leads to several intricacies not readily resolved by the use of "deterministic" concepts and models, as the work of Suijs and Borm (1999), Suijs, Borm, Wagenaere, and Tijs (1999), Blankenburg, Klusch, and Shehory (2003), Chalkiadakis and Boutilier (2004) and Chalkiadakis, Markakis, and Boutilier (2007) demonstrates. On a related note, enriching our model description so as to capture type uncertainty (Chalkiadakis & Boutilier, 2004; Chalkiadakis et al., 2007) would allow for the ready translation of uncertainty regarding the types (capabilities) of players to coalitional value uncertainty, while still capturing the potential stochasticity of coalitional action outcomes at the same time.

## 11. Acknowledgments

We would like to thank the anonymous reviewers for their constructive comments. This research was supported by the ALADDIN (Autonomous Learning Agents for Decentralised Data and Information Networks) project, which is jointly funded by a BAE Systems and EPSRC strategic partnership (EP/C548051/1); as well as by EPSRC (GR/T10664/01), ESRC (ES/F035845/1), and Singapore NRF Research Fellowship 2009-08.

## Appendix A. Algorithmic Aspects of Social Welfare Maximization in TTGs

In this appendix, we study the complexity of finding a social welfare-maximizing outcome in TTGs, both in the overlapping and in the non-overlapping scenario. Unless explicitly mentioned otherwise, we make the standard assumption that all parameters in the description of a TTG (i.e., all agents' weights, all thresholds and all task utilities), are integers given in binary.

It is not hard to see that finding a non-overlapping coalition structure that maximizes the social welfare is an NP-hard problem.

**Proposition 8.** *Given a TTG $G = (N; \boldsymbol{w}; \boldsymbol{t})$ and a parameter $K$, it is NP-complete to decide if $\hat{G}$ has an outcome $(CS, \boldsymbol{p})$ with $v(CS) \geq K$. This holds even if there is just one task type, i.e., $\boldsymbol{t} = t^1$, and all weights, thresholds and utilities are given in unary.*

*Proof.* It is easy to see that the problem is in NP. To show NP-hardness, we give a reduction from 3-PARTITION (Garey & Johnson, 1990) to our problem. An instance of 3-PARTITION is given by a list of non-negative integers $A = (a_1, \ldots, a_{3\ell})$ and an integer parameter $B$ that satisfies $\sum_{i=1}^{3\ell} = \ell B$ and $B/4 < a_i < B/2$ for all $i = 1, \ldots, 3\ell$. It is a "yes"-instance if the elements of $A$ can be partitioned into $\ell$ sets $S_1, \ldots, S_\ell$ such that $a(S_1) = \cdots = a(S_\ell) = B$ and a "no"-instance otherwise.

Given an instance of 3-PARTITION, consider a TTG $G$ with $N = \{1, \ldots, 3\ell\}$, $w_i = a_i$ for $i = 1, \ldots, 3\ell$ and a single task type $t = (T, u)$ with $T = B$ and $u = 1$. Clearly, deciding whether the maximum social welfare achievable in $\hat{G}$ is at least $\ell$ is equivalent to checking whether the given instance of 3-PARTITION is a "yes"-instance. Moreover, since 3-PARTITION is known to remain NP-hard when the input is given in unary, the same is true for our problem. □

In contrast, finding a social welfare-maximizing coalition structure in the OCF game that corresponds to a TTG is a somewhat easier problem. Indeed, we can simply add together all agents' weights, and then find an optimal set of tasks to execute given this amount of resource. The latter problem is equivalent to UNBOUNDED KNAPSACK, which is known to be NP-hard when the inputs are given in binary, but is polynomial-time solvable if all elements of the input are given in unary or if there are at most 2 items; for details, see (Martello & Toth, 1990), Section 3.6. Consequently, a similar conclusion holds for our problem.

**Proposition 9.** *Given a TTG $G = (N; \boldsymbol{w}; \boldsymbol{t})$ and a parameter $K$, it is NP-complete to decide if $\check{G}$ has an outcome $(CS, \boldsymbol{x})$ with $v(CS) \geq K$. However, this problem becomes polynomial-time solvable if all weights, thresholds and utilities are given in unary or if there are at most 2 task types.*